\documentclass[useAMS,usenatbib]{mn2e}
\usepackage{graphicx}
\usepackage{amssymb}

\usepackage{natbib,times}
\usepackage{lscape}

\usepackage{txfonts}

\newcommand{\mc}{\multicolumn}
\newcommand{\beqn}{\begin{equation}}
\newcommand{\eeqn}{\end{equation}}

\newcommand{\Hb}{{\rm H}$\beta$}
\newcommand{\Ha}{{\rm H}$\alpha$}

\newcommand{\Hdelta}{{\rm H}\delta}

\def\starlight{\textsc{starlight}}

\begin{document}

\title[Star formation activities in early-type BCGs]
{Star formation activities in early-type brightest cluster galaxies}
\author[F. S. Liu et al.]{
F. S. Liu$^{1,2}$\thanks{E-mail: lfs@nao.cas.cn},
Shude Mao$^{2, 3}$, 
X. M. Meng$^{2}$
\\
$^{1}$College of Physical Science and Technology,
Shenyang Normal University, Shenyang, 110034, P.R.China\\
$^{2}$National Astronomical Observatories, Chinese
Academy of Sciences, A20 Datun Road, Beijing, 100012, P.R.China\\
$^{3}$Jodrell Bank Centre for Astrophysics, University of Manchester, Manchester M13 9PL, UK\\
}

\date{Accepted 2012 March 8. Received 2012 March 7; in original form 2011 July 28}

\pagerange{\pageref{firstpage}--\pageref{lastpage}} \pubyear{2012}

\maketitle

\label{firstpage}

\begin{abstract}
We identify a total of 120 early-type Brightest Cluster Galaxies (BCGs) at $0.1<z<0.4$ in two recent large 
cluster catalogues selected from the Sloan Digital Sky Survey (SDSS). 
They are selected with strong emission lines in their optical spectra,  with both \Ha~and [O II]${\lambda}$3727 
line emission, which indicates significant ongoing star formation. 
They constitute about $\sim 0.5\%$ of the largest, optically-selected, low-redshift BCG sample, and 
the fraction is a strong function of cluster richness. Their star formation history can be well described by a recent minor and short starburst 
superimposed on an old stellar component, with the recent episode of star formation contributing on average only less than 1 percent of the total stellar mass. 
We show that the more massive star-forming BCGs in richer clusters tend to have higher star formation rate (SFR) 
and specific SFR (SFR per unit galaxy stellar mass).
We also compare their statistical properties with a control sample selected from X-ray luminous clusters, 
and show that the fraction of star-forming BCGs in X-ray luminous clusters is almost one order of magnitude 
larger than that in optically-selected clusters. BCGs with star formation in cooling flow clusters usually have 
very flat optical spectra and show the most active star formation, which may be connected with cooling flows. 
\end{abstract}


\section{Introduction}

The activity of star formation (SF) in a present-day galaxy is
strongly related to the local galaxy density and stellar mass 
\citep[][]{kwh+04}. Massive early-type galaxies lie in
higher density environments \citep[][]{Dressler80} and are dominated by
redder, older stars than less massive ones. The specific star
formation rate (i.e., SFR per unit stellar mass) of galaxies tends
to be lower in denser environments \citep[][]{kwh+04}, pointing to a picture where 
more massive galaxies form stars at a lower
rate per unit mass than less massive ones. Therefore, the
bulk of stars  in present-day massive galaxies must have formed at earlier
epochs than stars in less massive galaxies \citep[e.g.,][]{khw+03b,tmb+05}. The standard models of galaxy
formation have difficulty reproducing these red and dead massive
galaxies, unless feedback mechanisms (e.g., by active galactic nuclei-AGN) are introduced that prevent the gas
from cooling and forming stars. The star formation history 
of massive galaxies is not yet fully understood.

The Brightest Cluster Galaxies (BCGs) are at the most luminous and
massive end of galaxy population. They are usually located at or
close to the centres of dense clusters of galaxies
\citep[e.g.,][]{jf84,sks+05}. Most of them are dominated by old
stars without prominent ongoing star formation. It has been shown
that BCGs are different from other massive galaxies (non-BCGs) in
the surface brightness profiles and some basic scaling relations
\citep[e.g.,][]{mms64,Oemler73,Oemler76,Schombert86,Schombert87,Schombert88,glc+96,pmp+06,lfr+07,bhs+07,vbk+07,dqm+07,lxm+08},
which may indicate a distinct formation mechanism.

Recent studies from numerical simulations and semi-analytic models
in the cold dark matter hierarchical structure formation framework
indicate that a large part of stellar mass in BCGs may have formed
before redshift three, and later dry (dissipationless) mergers
play an important role in their stellar mass assembly
\citep{glp+04,db07}. This picture is largely consistent with
observations. For examples, many examples of dry mergers involving
central galaxies in groups and clusters at z$<$1 have been
reported \citep[e.g.,][]{lauer88,vff+99,mlf+06,jml+07,tvf+05,tmg+08,rfv07,mgh+08,lmd+09},
although some studies of BCGs in the more distant universe disagree with
this scenario \citep[e.g.,][]{wad+08,scb+11}.

The inclusion of AGN feedback in \citet{db07} can efficiently 
truncate the initial starburst and ensure that the progenitor of BCGs experiences virtually no star formation
in any evolution. However, some recent studies from the ultraviolet luminosities, infrared emission or line emission 
show increasing evidence for ongoing star formation and post-starbursts in some BCGs
\citep[e.g.,][]{allen95,cga98,cae+99,edge01,hm05,mrb+06,emr+06,wes06,ehb+07,obp+08,cdv+08,pkb+09,oqo+10,lwh+12}.
The existence of blue cores and UV excess in some BCGs are also interpreted as evidence for ongoing 
star formation \citep[e.g.,][]{bhb+08,pkb+09,wok+10,hmd10}. Although  active star formation in these BCGs 
is compelling, the starbursts may have very short timescales (shorter than 200 Myr) and only contribute a small mass fraction (less than 1 percent, \citealt{pkb+09}).      

The BCGs with ongoing SF studied in previous works are mostly selected 
from X-ray cluster samples, and  usually reside in cooling flow clusters.  
It has been shown that star formation in these BCGs is correlated with the cooling timescale ($\rm t_{cool}$) of the gas \citep[][]{rmn08}, which is a strong 
indicator of their connections. However, previous studies are based on small samples and are biased 
toward X-ray luminous clusters, which may not be a representative of this population.  
It has also been shown nearby optically-selected local BCGs (e.g., z$<$0.1) have little indication 
for enhanced active star formation \citep{ehb+07,vbk+07,wok+10}. In this study, we search for BCGs with ongoing 
SF in clusters at higher redshift. We select a sample of 120 early-type BCGs at $0.1<z<0.4$ from two large 
optically-selected cluster catalogues of SDSS-WHL \citep[][]{WHL09} and GMBCG \citep[][]{hmk+10}. 
This sample is roughly an order of magnitude larger than previous ones. 
They are selected with strong emission lines in their optical spectra, with both \Ha~and [O II]${\lambda}$3727 
line emission, which indicates significant ongoing star formation. 
We investigate their statistical properties and make a comparison with 
a control sample selected from X-ray luminous clusters. For the first time, we probe the dependence 
of SF activities in these BCGs on their stellar masses and cluster environments. 
We also reconstruct their star formation history using stellar population synthesis models, 
and discuss their physical connections with the cooling flows and galactic cannibalism.

The structure of the paper is as follows. We describe our sample selection and data analysis in \S2 and \S3, and
present our results in \S4.  A summary  and discussion are given in \S5. 
Throughout this paper we adopt a cosmology with a matter density parameter
$\Omega_{\rm m}=0.3$, a cosmological constant
$\Omega_{\rm\Lambda}= 0.7$, and a Hubble constant of $H_0=70\,{\rm
km \, s^{-1} Mpc^{-1}}$, i.e., $h=H_0/(100\,{\rm km \,  s^{-1} Mpc^{-1}})=0.7$.

\section{Sample Selection\label{sec:sample}}

We identify early-type BCGs with significant star formation from two large optically-selected 
cluster catalogues of SDSS-WHL \citep[][]{WHL09} and GMBCG \citep[][]{hmk+10}.  
The SDSS-WHL cluster catalogue was constructed from the SDSS DR6 photometric galaxy catalogue,
 which includes 39,668 clusters in the redshift 
range $0.05<z<\sim0.6$ with more than eight luminous ($M_r\leq-21$) member galaxies 
within a radius of $0.5$\,Mpc and a photometric redshift interval $z\pm0.04(1+z)$. 
The GMBCG catalogue was constructed from the SDSS DR7 photometric catalogue by identifying the red sequence plus BCG feature, 
which includes 55,424 clusters in the redshift range $0.1<z<0.55$. We use the data at $0.1<z<0.4$ 
because both catalogues are relatively complete out to $z\sim0.4$. There is another cluster catalogue by \citet[][]{spd+11}. 
we do not use this sample since it is not yet public at the time of writing. These three cluster samples overlap 
but also differ in their lists of clusters. SDSS-WHL and GMBCG samples give fully consistent results (see below) concerning SF activities, and thus our conclusions should not be much affected by which catalogue we use.

In our selection, we require BCGs to have spectroscopic observations and their spectra have been parameterised by the MPA/JHU
team\footnote{http://www.mpa-garching.mpg.de/SDSS/DR7/}. We discard the BCGs with concentration index $C=R_{90}/R_{50}<2.5$ in the $i-$band (to select 
early-type objects, \citealt[][]{bsa+03}) and those with a median signal-to-noise ratio (S/N) per pixel of the whole spectrum smaller than 3. 
As a result, we obtain 10,996 and 15,181 early-type BCGs at $0.1<z<0.4$ from the SDSS-WHL and GMBCG catalogues, respectively. 

\begin{figure*}
\centering
\includegraphics[angle=0,width=0.7\textwidth]{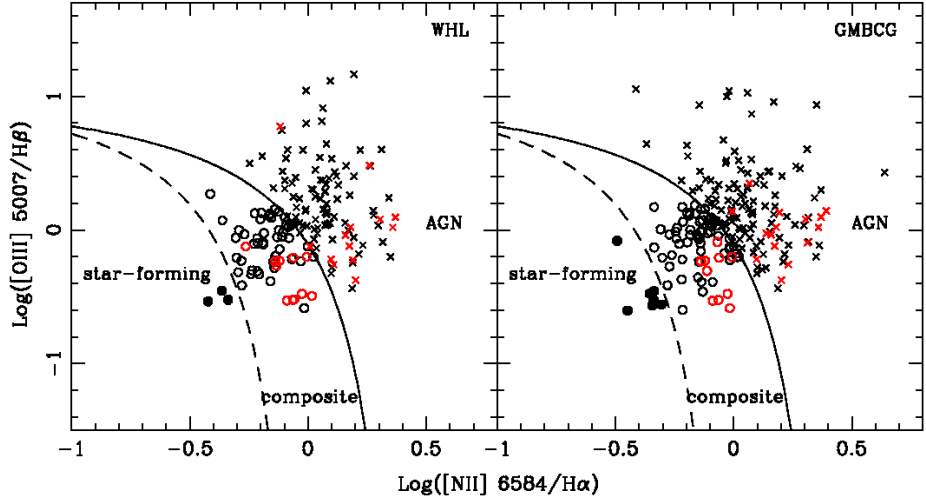}
\caption{The diagnostic diagrams of [NII]$\lambda$6584/{\Ha} versus [OIII]$\lambda$5007/{\Hb} for
emission-line early-type BCGs in the SDSS-WHL catalogue (left panel) and GMBCG catalogue (right panel), respectively. 
The purely star-forming galaxies, composites and AGNs are shown with solid circles, open circles and crosses, respectively.
The sources in X-ray luminous clusters are shown with red symbols. 
The solid line is from \citet{kgk+06}, and the dashed line is from \citet{kht+03}.
} \label{fig:diagnose}
\end{figure*}

\citet{bpt81} proposed a suite of three diagnostic diagrams to classify the dominant energy source in emission-line galaxies, 
which are commonly known as the Baldwin-Phillips-Terlevich (BPT) diagrams and are based on four emission line ratios, 
[OIII]$\lambda$5007/\Hb, [NII]$\lambda$6584/\Ha, [SII]$\lambda$6716+6731/{\Ha} and [OI]$\lambda$6300/\Ha. 
Here, we only use the emission line diagnostic diagram of [OIII]$\lambda$5007/\Hb~versus [NII]$\lambda$6584/{\Ha} 
because the other lines ([SII]$\lambda$6716,6731 and [OI]$\lambda$6300) 
are usually very weak in early-type galaxies \citep{hg09}. 
We select the emission-line BCGs, requiring 
(1) the lines of [NII]$\lambda$6584, H$\alpha$, [OIII]$\lambda$5007, H$\beta$ and [O II]$\lambda$3727 are detected as 
emission lines and have the S/N$>$3. 
(2) the equivalent widths (EWs) of both H$\alpha$~and [O II]$\lambda$3727 lines are greater than 3$\AA$.
It should be noted that some previous studies (e.g., Crawford et al. 1999; Donahue et al. 2010) have shown that 
the line emission in BCGs with large equivalent width of H$\alpha$~is dominated by star formation. The cut of 
H$\alpha$~EW $>3\AA$ is thus strongest in our criteria to select sources with SF activities. The cut of [O II] EW $>3\AA$ 
only reject $\sim1\%$ of sources with large H$\alpha$~EW, which does not affect our statistical result. 
%
However, the inclusion of the [O II]$\lambda$3727 line here allows us to investigate the origin of [O II]$\lambda$3727 line emission in 
star-forming BCGs (see \S3.1).
%
Our criteria inevitably reject weak emission-line BCGs. It is acceptable since the ability to distingush their types by the BPT 
diagram will be poor \citep[e.g.,][]{hst+05}.   
In total, 159 and 201 objects satisfy our criteria in the selected SDSS-WHL and GMBCG early-type BCGs respectively. 
The fraction is $\sim1.4\%$ (159/10,996) for SDSS-WHL sample, $\sim1.3\%$ (201/15,181) for GMBCG sample, respectively.

The diagnostic diagram mentioned above then classifies these emission-line BCGs into different types, which are shown in Figure~\ref{fig:diagnose}. 
We here detect 3 purely star-forming objects, 59 composites (SF $+$ AGN) and 98 AGNs in the SDSS-WHL objects (see the left panel of Figure~\ref{fig:diagnose}), 
and 7 purely star-formings objects, 71 composites and 123 AGNs in the GMBCG objects (right panel).   
We select those classified as purely star-forming objects or composites as our targets, which constitute a very rare population relative 
to the whole sample ($62/10996\sim0.56\%$ for SDSS-WHL objects, $78/15181\sim0.5\%$ for GMBCG objects). Notice that 20 of these targets 
overlap in these two catalogues. Thus we finally obtain a total of 120 early-type BCGs with significant ongoing star formation  
(9 purely star-forming objects and 111 composites), which are listed in Table 1.

We also identify a control BCG sample from X-ray luminous clusters. The {\it ROSAT} All Sky Survey detected 18,806 
bright sources \citep{vab+99} and 105,924 faint sources \citep{vab+00} in the
0.1--2.4 keV band, of which 378 extended sources in the northern hemisphere 
and 447 extended sources in the southern hemisphere have been identified as clusters of galaxies  \citep{bvh+00,bsg+04}. 
However, many objects in recent catalogues of SDSS clusters may also be unidentified X-ray clusters \citep[][]{WHL09} since 
SDSS have detected many new clusters of galaxies. We follow \citet[][]{WHL09} to cross-identify the {\it ROSAT} 
X-ray bright and faint sources with two spectroscopic catalogues to construct a new sample of X-ray cluster candidates. 
We first select X-ray sources with a projected separation of $r_p<0.3$ Mpc from the BCGs and hardness ratios of 0--1  as our targets \citep[][]{WHL09}. 
In total, we obtain 112 targets in the {\it ROSAT} bright source catalog and 194 targets in the {\it ROSAT} faint source catalog for SDSS-WHL clusters, respectively.   
We also obtain 125 targets in the {\it ROSAT} bright source catalog and 236 targets in the {\it ROSAT} faint source catalog for GMBCG clusters, respectively.

\begin{figure}
\centering
\includegraphics[angle=0,width=0.47\textwidth]{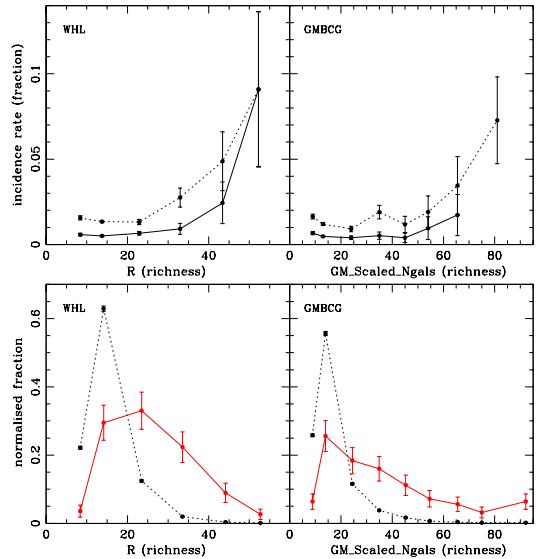}
\caption{Top panels: the incidence rates (fractions) of identified emission-line BCGs (dots connected with dashed line)
and BCGs with SF (dots connected with solid line) as a function of cluster richness for SDSS-WHL clusters (left) and GMBCG clusters (right), respectively.
Bottom panels: the normalised distributions of cluster richness of optically-selected samples (dots connected with dashed line)
and X-ray luminous sample (dots connected with solid line) for SDSS-WHL clusters (left) and GMBCG clusters (right), respectively.
Poisson errors are shown.
} \label{fig:fraction}
\end{figure}

We first cross-correlate selected emission-line BCGs with these X-ray candidates.
There are 24 SDSS-WHL emission-line BCGs and 25 GMBCG emission-line BCGs in {\it ROSAT} bright source catalog respectively.  
Only 3 emission-line BCGs and 5 GMBCG emission-line BCGs are found in the {\it ROSAT} faint source catalog.
The fractions of emission-line BCGs relative to the X-ray luminous samples ($24/112\sim21\%$ for SDSS-WHL objects, $25/125 \sim 20\%$ for GMBCG objects) 
are almost one order of magnitude higher than that in optically-selected sample ($\sim1.4\%$). 
The derived fraction in X-ray luminous sample ($\sim20\%$) is slightly lower than previous 
results (e.g., $27\%$ from Crawford et al. 1999; $22\%$ from Donahue et al. 2010). The difference may be a result of 
different selection criteria or sample sizes. \citet[][]{dbw+10} analysed a small sample (with 32 objects), and the detected emission-line BCGs 
by H$\alpha$~EW $>1\AA$ showed the typical forbidden line emission as well. Our incidence rate is close to their result. 
\citet[][]{cae+99} used a relatively large statistical sample (with 256 central dominant galaxies) and selected emission-line galaxies based on H$\alpha$~line emission only. 
If we select emission-line BCGs by H$\alpha$~emission line only and apply the same H$\alpha$~luminosity detection limits as \citet[][]{cae+99}, 
namely extrapolating their slope to our redshift range according to the expected $L \propto z^2$ relation (see Crawford et al. 1999 for details) 
and correcting the difference in cosmological parameters. As a result, We obtain 29 SDSS-WHL H$\alpha$ emitters and 35 GMBCG H$\alpha$ emitters 
in X-ray luminous clusters. The fractions ($29/112\sim26\%$ for SDSS-WHL objects, $35/125 \sim 28\%$ for GMBCG objects) are almost the same 
with that of \citet[][]{cae+99}.

We then cross-correlate those BCGs with SF with these X-ray candidates. There are 11 SDSS-WHL BCGs with SF and 10 GMBCG BCGs with SF 
in {\it ROSAT} bright source catalog. Notice that 8 sources with SF overlap in these two X-ray samples. Therefore 13 out of a total of 120
early-type BCGs with SF are likely to be in X-ray luminous clusters. In fact, 11 of these 13 sources are  known
X-ray luminous clusters according to the NASA/IPAC Extragalactic Database (NED), which have been indicated in Table 1. 
The fractions of BCGs with SF relative to the whole X-ray luminous
samples ($11/112\sim9.8\%$ for SDSS-WHL objects, $10/125 \sim 8.0\%$ for GMBCG objects) are also one order of magnitude higher than
that in optically-selected sample ($\sim 0.5\%$). 
We show the fractions of these BCGs with SF (solid line) and selected emission-line BCGs (dashed line) as
a function of cluster richenss for the SDSS-WHL objects (top-left panel) and GMBCG objects (top-right panel) in Figure~\ref{fig:fraction} respectively.
Notice that the relations with cluster richness for sources in these two catalogues 
are shown separately because the cluster richness in these catalogues is estimated with different algorithms.
It can be seen that the more massive clusters tend to habour higher fractions of emission-line BCGs and SF BCGs. 
The fractions are usually the highest in the richest clusters. It indicates that the incidence rates of emission-line BCGs and SF BCGs 
in a cluster sample may be much higher above some minimum cluster richness (mass). It can thus be understood that the incidence rates of 
emission-line BCGs and BCGs with SF are higher in X-ray luminous clusters than optically-selected ones since X-ray selected clusters 
are usually more massive (see bottom panels of Figure~\ref{fig:fraction}).

\begin{table*}
\scriptsize
\begin{minipage}{175mm}
\setlength{\tabcolsep}{0.025in}
\caption[]{Basic parameters for the 120 identified early-type BCGs with ongoing star formation.}
\begin{center}
\begin{tabular}{lcrrccccccccccccc}
\hline
\mc{1}{c}{Cluster Name } &
\mc{1}{c}{ X-ray Cluster } & 
\mc{1}{c}{BCG R.A. } & 
\mc{1}{c}{BCG Dec. } &
\mc{1}{c}{BCG $z$ } & 
\mc{1}{c}{${\rm log M_{\ast,fib}}$ } & 
\mc{1}{c}{${\rm log M_{\ast,tot}}$ } &
\mc{1}{r}{$f_{[\rm N II]\lambda6584 }$ } &
\mc{1}{c}{$f_{\rm H_{\alpha}}$ } &    
\mc{1}{r}{$f_{[\rm O III]\lambda5007 }$ } &
\mc{1}{c}{$f_{\rm H_{\beta}}$ } &
\mc{1}{r}{$f_{[\rm O II]\lambda3727}$ } & 
\mc{1}{c}{${\rm SFR({H_{\alpha}})}$ } &
\mc{1}{c}{${\rm SFR({[\rm O II]})}$ } &
\mc{1}{c}{${\rm D_n(4000)}$ } &
\mc{1}{c}{${\rm H{\delta}_A}$ } &
\mc{1}{c}{$\rm Type$ } \\

\mc{1}{c}{(1)} & \mc{1}{c}{(2)} & \mc{1}{c}{(3)} & \mc{1}{c}{(4)} & \mc{1}{c}{(5)} & \mc{1}{c}{(6)} & \mc{1}{c}{(7)} & \mc{1}{c}{(8)} & \mc{1}{c}{(9)} & \mc{1}{c}{(10)} & \mc{1}{c}{(11)} & \mc{1}{c}{(12)} & \mc{1}{c}{(13)} & \mc{1}{c}{(14)} & \mc{1}{c}{(15)} & \mc{1}{c}{(16)} & \mc{1}{c}{(17)} \\
\hline
WHLJ004721.3+005239& &11.83866& 0.87768& 0.1025& 10.66& 11.25& 241.18& 397.26& 118.32& 145.05& 372.28& 0.84& 0.56& 1.72& -0.88& comp \\
WHLJ015435.9+002641& &28.64954& 0.44486& 0.2288& 10.81& 11.41& 31.24& 50.85& 14.11& 17.74& 78.05& 0.62& 0.73&  1.93& -3.21& comp \\
WHLJ024253.6$-$065742$^{a}$& &40.72351& -6.96174& 0.3503& 11.08& 11.67& 27.80& 32.63& 12.34& 11.38& 35.41& 1.07& 0.85&  1.75& -4.46& comp \\
WHLJ080641.4+494628& &121.69920& 49.79065& 0.2434& 10.89& 11.65& 149.48& 199.03& 68.60& 71.28& 432.16& 2.81& 5.01&  1.58& -1.00& comp \\
WHLJ081728.1+065903$^{a}$& &124.36710& 6.98433& 0.2565& 10.88& 11.47& 20.96& 45.57& 6.16& 20.57& 41.50& 0.73& 0.50&  1.85& -1.82& sb \\
WHLJ085411.2+190702$^{a}$& &133.54660& 19.11724& 0.1813& 10.94& 11.53& 181.32& 263.25& 119.93& 95.83& 652.58& 1.92& 3.86&  1.85& -3.22& comp \\
WHLJ090940.6+105005& &137.41920& 10.83483& 0.1402& 10.79& 11.31& 947.18& 1873.65& 635.54& 718.86& 8240.88&  7.77& 29.11&  1.45& 3.12& comp \\
WHLJ092018.8+370618$^{a}$& &140.07820& 37.10510& 0.2348& 10.79& 11.55& 221.31& 222.38& 57.69& 80.22& 422.54& 2.90& 4.52&  1.78& 0.75& comp \\
WHLJ092243.7+351448& &140.68201& 35.24673& 0.2307& 10.97& 11.46& 701.67& 1390.30& 230.58& 510.79& 1141.88& 17.40& 11.87& 1.41& 3.96& comp \\
WHLJ092609.4+670407& &141.53909& 67.06886& 0.1211& 10.97& 11.40& 1847.34& 3034.38& 512.55& 1116.11& 2684.82& 9.17& 6.67&  1.41& 1.85& comp \\
WHLJ092954.3+002752& &142.47610& 0.46474& 0.1457& 10.51& 11.24& 60.35& 82.82& 18.35& 29.19& 58.32& 0.37& 0.17& 1.69& 1.31& comp \\
WHLJ101652.1+135938& &154.21809& 13.97771& 0.1455& 10.54& 11.15& 137.06& 239.31& 38.82& 85.37& 335.11& 1.08& 1.20& 1.52& -0.75& comp \\
WHLJ102339.6+041110$^{a,b}$& Z3146&155.91521& 4.18628& 0.2897& 11.06& 11.73& 4650.55& 6172.13& 1251.03& 2245.14& 10896.93& 129.85& 196.85&   1.18& 4.94& comp \\
WHLJ102831.8+151511$^{a}$& &157.13229& 15.25325& 0.3046& 10.92& 11.59& 143.72& 136.80& 29.65& 49.20& 212.56& 3.23& 4.11& 1.76& 0.14& comp \\
WHLJ103545.5+152435& &158.93961& 15.40980& 0.2582& 11.11& 11.57& 116.78& 144.83& 54.63& 51.79& 340.51& 2.34& 4.50&   1.89& 0.72& comp \\
WHLJ104949.8+054629& &162.45770& 5.77494& 0.2640& 11.10& 11.74& 279.52& 378.97& 115.06& 139.19& 1308.80& 6.44& 18.84&  1.86& -0.21& comp \\ 
WHLJ105158.8+082222& &162.99510& 8.37297& 0.1884& 10.94& 11.36& 108.95& 131.29& 39.50& 46.91& 227.86& 1.04& 1.42&   1.79& -1.59& comp \\
WHLJ111320.5+173541$^{a,b}$& A1204&168.33540& 17.59474& 0.1705& 10.62& 11.39& 293.18& 298.27& 64.69& 107.25& 681.54& 1.90& 3.58&  1.73& -0.26& comp \\
WHLJ112154.6+305515& &170.47729& 30.92111& 0.2432& 10.88& 11.43& 37.67& 43.62& 16.78& 15.36& 51.62& 0.61& 0.52&    1.65& 1.47& comp \\
WHLJ112536.8+592155& &171.40331& 59.36544& 0.3104& 11.03& 11.32& 762.28& 1100.91& 156.79& 396.99& 615.92& 27.15& 12.57&  1.21& 3.58& comp \\
WHLJ112612.8$-$005130& &171.55341& -0.85854& 0.2553& 11.10& 11.73& 116.73& 136.70& 28.73& 48.65& 221.30& 2.15& 2.83&  1.79& -0.24& comp \\
WHLJ112714.9+482220& &171.81219& 48.37249& 0.1679& 10.82& 11.35& 106.22& 244.62& 99.86& 86.91& 293.73& 1.51& 1.37&  1.57& 0.80& comp \\
WHLJ113520.7+491127& &173.83611& 49.19104& 0.1314& 10.61& 11.11& 147.34& 254.67& 43.64& 92.06& 265.07& 0.92& 0.73& 1.48& 0.82& comp \\
WHLJ113957.1+681118& &174.87270& 68.17191& 0.1543& 10.84& 11.45& 647.91& 899.22& 184.26& 327.60& 1323.14& 4.60& 5.58&  1.51& 2.84& comp \\
WHLJ121256.2+272657$^{a,b}$& &183.24640& 27.45126& 0.1797& 10.37& 11.37& 470.91& 489.82& 44.03& 176.00& 461.33& 3.50& 2.70&  1.75& -0.70& comp \\
WHLJ122205.0$-$013609& &185.52090& -1.60274& 0.2005& 10.85& 11.55& 342.95& 493.45& 208.99& 181.58& 1141.12& 4.50& 8.62&   1.66& -1.04& comp \\
WHLJ125230.0+035803& &193.12500& 3.96769& 0.1942& 10.90& 11.42& 249.15& 479.29& 97.08& 174.67& 566.97& 4.08& 3.92&  1.54& 2.26& comp \\
WHLJ131330.4+320039$^{a}$& &198.38640& 32.03464& 0.3042& 10.98& 11.64& 34.19& 53.93& 25.10& 19.19& 148.49& 1.27& 2.82&   1.92& -0.48& comp \\
WHLJ131451.7+383418$^{a}$& &198.71539& 38.57187& 0.2360& 10.89& 11.53& 67.15& 92.61& 45.78& 32.96& 234.94& 1.22& 2.47&   1.99& -1.53& comp \\
WHLJ132414.6+041803$^{b}$& RX J1324.2+0419 &201.08200& 4.31862& 0.2631& 10.62& 11.48& 233.38& 224.90& 24.91& 80.19& 380.76& 3.79& 5.38& 1.69& 0.68& comp \\
WHLJ132858.3$-$005343& &202.27600& -0.89491& 0.2373& 11.00& 11.52& 55.89& 71.10& 23.68& 25.16& 108.55& 0.95& 1.11&   1.63& 1.56& comp \\
WHLJ135321.5+395909$^{a}$& &208.33971& 39.98603& 0.1057& 10.75& 11.25& 48.23& 96.43& 20.78& 33.65& 123.41& 0.22& 0.19&  1.97& -1.68& comp \\
WHLJ135742.8+303505& &209.32860& 30.59559& 0.2085& 10.85& 11.31& 448.09& 1040.45& 125.38& 379.33& 920.91& 10.37& 7.62&  1.36& 3.50& sb \\
WHLJ140102.1+025242$^{a,b}$& A1835&210.25861& 2.87847& 0.2520& 11.18& 11.75& 2461.91& 3390.42& 679.54& 1226.92& 5268.39& 51.81& 68.71& 1.15& 5.42& comp \\
WHLJ141520.0+240036& &213.81760& 24.02097& 0.1386& 10.32& 11.19& 76.94& 87.80& 9.20& 30.70& 91.00& 0.36& 0.27&  1.90& -1.11& comp \\
WHLJ141835.5+020507$^{a}$& &214.64799& 2.08549& 0.2697& 10.94& 11.45& 280.08& 394.97& 78.62& 145.19& 412.03& 7.05& 6.03&   1.38& 1.88& comp \\
WHLJ141837.7+374624$^{a}$& &214.65691& 37.77348& 0.1349& 10.68& 11.33& 76.26& 129.25& 35.46& 45.15& 160.96& 0.49& 0.44&  1.90& -2.54& comp \\
WHLJ142355.5+262623$^{a,b}$& RX J1423.9+2626 &215.98109& 26.43979& 0.1482& 10.56& 11.36& 345.25& 399.16& 85.42& 143.30& 789.69& 1.87& 3.03& 1.64& 0.38& comp \\
WHLJ142424.3+251427& &216.10139& 25.24108& 0.2331& 10.63& 11.39& 351.65& 378.22& 76.03& 136.26& 520.47& 4.85& 5.49&  1.44& 3.34& comp \\
WHLJ143340.9$-$014503& &218.42059& -1.75099& 0.2194& 10.90& 11.52& 90.88& 119.79& 29.76& 42.57& 97.48&  1.34& 0.80&  1.85& -2.19& comp \\
WHLJ144048.8+150625& &220.14830& 15.13004& 0.1141& 10.41& 11.04& 890.17& 2365.53& 237.56& 865.68& 1697.03& 6.28& 3.68&  1.31& 3.41& sb \\
WHLJ144621.2+381525& &221.58850& 38.25720& 0.2344& 10.81& 11.31& 417.79& 849.85& 253.57& 316.54& 717.50&  11.03& 7.54&  1.45& 3.77& comp \\
WHLJ145213.4+053857& &223.05569& 5.64942& 0.2303& 11.02& 11.49& 147.02& 195.47& 92.66& 69.86& 273.04& 2.44& 2.66&  1.70& -0.65& comp \\
WHLJ145715.1+222034$^{a,b}$& MS 1455.0+2232&224.31300& 22.34288& 0.2576& 10.96& 11.68& 517.78& 601.64& 62.33& 215.92& 1166.08& 9.67& 15.94&   1.55& 3.46& comp \\
WHLJ150407.5$-$024816$^{b}$& RXC J1504.1-0248&226.03130& -2.80460& 0.2169& 11.06& 11.57& 4239.69& 7795.80& 2027.40& 2788.25& 13462.17& 84.91& 125.45&   1.16& 4.98& comp \\
WHLJ151303.8+252550& &228.26570& 25.43079& 0.1835& 10.76& 11.24& 267.15& 427.17& 177.00& 157.21& 259.42& 3.20& 1.40&  1.60& 0.85& comp \\
WHLJ153253.8+302059$^{b}$& RX J1532.8+3021&233.22411& 30.34983& 0.3620& 10.99& 11.57& 2090.19& 2843.55& 531.97& 1022.70& 4851.62& 100.34& 146.85&   1.18& 6.11& comp \\
WHLJ153915.3+422950& &234.81390& 42.49735& 0.2335& 10.99& 11.49& 99.26& 120.47& 43.95& 42.70& 69.06& 1.55& 0.60& 1.56& 1.67& comp \\
WHLJ154606.6+120650$^{a}$& &236.52740& 12.11406& 0.1849& 10.95& 11.51& 137.32& 172.02& 73.83& 61.90& 522.95& 1.31& 3.25&  2.05& -2.97& comp \\
WHLJ161228.3+113547& &243.09700& 11.59213& 0.2716& 11.19& 11.77& 1582.66& 2447.76& 790.20& 916.63& 5834.31& 44.40& 90.44& 1.34& -1.32& comp \\
WHLJ161621.7+441914& &244.09019& 44.32083& 0.1951& 10.93& 11.39& 149.92& 233.20& 64.58& 85.39& 218.13& 2.00& 1.43& 1.70& 0.10& comp \\
WHLJ162044.2+125214& &245.18410& 12.87080& 0.1904& 10.96& 11.42& 62.21& 90.90& 40.16& 32.19& 177.38& 0.74& 1.10&  1.73& -3.11& comp \\
WHLJ162302.1+475939& &245.75150& 47.96708& 0.1993& 10.83& 11.37& 207.54& 322.69& 86.34& 119.60& 582.09& 2.91& 4.29&  1.57& 1.19& comp \\
WHLJ162309.4+440441& &245.78931& 44.07832& 0.1333& 10.89& 11.57& 1097.92& 1853.04& 844.61& 689.20& 4452.19& 6.89& 13.81& 1.70& 0.14& comp \\
WHLJ163936.4+370501$^{a}$& &249.90150& 37.08373& 0.1826& 10.57& 11.15& 262.64& 501.06& 66.52& 180.96& 363.31& 3.71& 2.14&   1.38& 1.91& comp \\
WHLJ170046.3+222141& &255.19141& 22.39863& 0.2003& 11.11& 11.54& 368.65& 958.26& 610.92& 354.42& 2033.18& 8.73& 15.33&   1.65& 0.43& comp \\
WHLJ172010.0+263732$^{a,b}$& RX J1720.2+2637&260.04181& 26.62557& 0.1601& 10.40& 11.40& 537.32& 659.43& 67.68& 234.29& 860.97& 3.66& 3.97&  1.57& 1.71& comp \\
WHLJ172423.0+273242& &261.09579& 27.54511& 0.2330& 10.87& 11.52& 326.10& 454.73& 81.77& 166.37& 1666.45& 5.82& 18.16&    1.83& -1.75& comp \\
WHLJ210101.3$-$070019& &315.25549& -7.00530& 0.1363& 10.84& 11.41& 74.76& 96.10& 29.69& 33.94& 197.98& 0.38& 0.59&   1.85& -1.43& comp \\
WHLJ212940.0+000521$^{a,b}$& RX J2129.6+0005 &322.41650& 0.08921& 0.2339& 10.38& 11.28& 218.52& 231.98& 26.75& 82.07& 332.73& 3.00& 3.56&   1.83& -0.38& comp \\
WHLJ215403.3+000950& &328.51370& 0.16403& 0.2144& 10.91& 11.40& 1019.89& 1905.20& 608.40& 702.94& 1310.39& 20.22& 11.30& 1.23& 5.66& comp \\
WHLJ232316.1+005922& &350.81711& 0.98978& 0.1092& 10.67& 11.34& 64.45& 103.68& 16.80& 36.49& 112.99& 0.25& 0.19&  1.94& -2.80& comp \\

\hline
\end{tabular}
\end{center}
\label{tab:bcg}
\end{minipage}
\end{table*}

\addtocounter{table}{-1}
\begin{table*}
\scriptsize
\begin{minipage}{175mm}
\setlength{\tabcolsep}{0.025in}
\caption[]{- continued}
\begin{center}
\begin{tabular}{lcrrccccccccccccc}
\hline
\mc{1}{c}{Cluster Name } &
\mc{1}{c}{X-ray Cluster } &
\mc{1}{c}{BCG R.A. } &
\mc{1}{c}{BCG Dec. } &
\mc{1}{c}{BCG $z$ } &
\mc{1}{c}{${\rm log M_{\ast,fib}}$ } &
\mc{1}{c}{${\rm log M_{\ast,tot}}$ } &
\mc{1}{r}{$f_{[\rm N II]\lambda6584 }$ } &
\mc{1}{c}{$f_{\rm H_{\alpha}}$ } &
\mc{1}{r}{$f_{[\rm O III]\lambda5007 }$ } &
\mc{1}{c}{$f_{\rm H_{\beta}}$ } &
\mc{1}{r}{$f_{[\rm O II]\lambda3727}$ } &
\mc{1}{c}{${\rm SFR({H_{\alpha}})}$ } &
\mc{1}{c}{${\rm SFR({[\rm O II]})}$ } &
\mc{1}{c}{${\rm D_n(4000)}$ } &
\mc{1}{c}{${\rm H{\delta}_A}$ } &
\mc{1}{c}{$\rm Type$ } \\

\mc{1}{c}{(1)} & \mc{1}{c}{(2)} & \mc{1}{c}{(3)} & \mc{1}{c}{(4)} & \mc{1}{c}{(5)} & \mc{1}{c}{(6)} & \mc{1}{c}{(7)} & \mc{1}{c}{(8)} & \mc{1}{c}{(9)} & \mc{1}{c}{(10)} & \mc{1}{c}{(11)} & \mc{1}{c}{(12)} & \mc{1}{c}{(13)} & \mc{1}{c}{(14)} & \mc{1}{c}{(15)} & \mc{1}{c}{(16)} & \mc{1}{c}{(17)} \\
\hline

GMBCGJ057.50891+00.06656& &57.50892& 0.06657& 0.1314& 10.92& 11.42& 68.05& 103.65& 33.25& 36.34& 124.35& 0.37& 0.31&  1.84& -2.05& comp \\
GMBCGJ121.87813+34.01156& &121.87810& 34.01156& 0.2079& 10.55& 11.50& 327.36& 318.73& 73.28& 115.70& 849.25& 3.16& 7.04&  1.80& -2.37& comp \\
GMBCGJ122.22965+46.81917& &122.22970& 46.81917& 0.1259& 10.73& 11.25& 445.18& 603.43& 71.65& 219.51& 452.93& 1.98& 1.16&  1.49& 1.96& comp \\
GMBCGJ129.88700+27.26509& &129.88699& 27.26509& 0.2839& 11.20& 11.68& 195.18& 286.61& 150.87& 105.91& 1308.43& 5.75& 22.22&   2.08& -3.40& comp \\
GMBCGJ130.54583+59.92378& &130.54581& 59.92378& 0.1278& 10.88& 11.35& 1734.54& 3852.76& 360.60& 1418.17& 1829.02& 13.08& 5.06&  1.44& 4.24& sb \\
GMBCGJ135.53378+37.26415& &135.53380& 37.26415& 0.2931& 11.10& 11.67& 214.91& 224.36& 57.81& 80.94& 247.33& 4.85& 4.32&    1.83& -1.53& comp \\
GMBCGJ138.55769+03.11440& &138.55769& 3.11440& 0.1420& 10.90& 11.29& 309.65& 582.08& 185.50& 212.30& 625.94& 2.48& 2.08&  1.38& 2.27& comp \\
GMBCGJ139.36943+07.70896& &139.36940& 7.70896& 0.1297& 10.75& 11.34& 53.06& 66.90& 26.49& 23.47& 121.61&  0.23& 0.30&  1.99& -2.59& comp \\
GMBCGJ140.61017+51.89892& &140.61020& 51.89892& 0.2016& 10.90& 11.48& 328.18& 451.00& 63.18& 165.38& 506.11& 4.17& 3.84&  1.59& 1.62& comp \\
GMBCGJ141.35050+17.33344& &141.35049& 17.33344& 0.2145& 10.90& 11.42& 61.79& 107.95& 32.28& 37.81& 136.72& 1.15& 1.11&  1.96& -0.49& comp \\
GMBCGJ141.68858+29.56935& &141.68860& 29.56935& 0.2026& 10.95& 11.42& 40.54& 42.19& 20.29& 28.28& 33.45& 0.39& 0.19&  1.77& -2.24& comp \\
GMBCGJ142.56414+25.22437& &142.56410& 25.22437& 0.3167& 11.31& 11.68& 70.43& 97.67& 32.70& 35.48& 205.58& 2.52& 4.34&  1.95& -1.62& comp \\
GMBCGJ145.15501+20.34164& &145.15500& 20.34164& 0.2474& 11.09& 11.59& 36.18& 44.61& 16.85& 15.71& 103.98& 0.65& 1.19&  1.92& -1.56& comp \\
GMBCGJ149.95279+12.63987& &149.95280& 12.63987& 0.2162& 10.85& 11.40& 322.32& 534.42& 271.16& 200.12& 2292.38& 5.78& 20.95&  1.74& 1.67& comp \\
GMBCGJ150.29590+33.96550& &150.29590& 33.96550& 0.1997& 10.90& 11.47& 54.25& 73.82& 25.03& 26.18& 139.25& 0.67& 0.97&  1.91& -3.13& comp \\
GMBCGJ150.82925+06.39045& &150.82930& 6.39045& 0.1841& 10.76& 11.37& 336.75& 516.35& 177.62& 191.77& 928.44& 3.90& 5.74&  1.78& -1.44& comp \\
GMBCGJ151.96927+27.50322& &151.96930& 27.50323& 0.1482& 11.01& 11.46& 1070.10& 2334.07& 280.03& 856.57& 1781.68& 10.93& 6.89&  1.46& 2.47& sb \\
GMBCGJ156.24967+39.94290& &156.24969& 39.94291& 0.3097& 11.10& 11.62& 31.49& 30.40& 7.47& 11.95& 55.66& 0.75& 1.07&  1.88& -1.11& comp \\
GMBCGJ156.85843+47.96615& &156.85851& 47.96611& 0.1317& 10.72& 11.42& 111.66& 158.28& 24.13& 55.77& 201.05& 0.57& 0.56&   1.96& -1.50& comp \\
GMBCGJ162.38906+24.15104& &162.38910& 24.15105& 0.2488& 11.00& 11.69& 211.09& 269.86& 109.21& 99.14& 920.85& 4.01& 11.46& 1.99& -1.47& comp \\
GMBCGJ162.59096+37.18131& &162.59100& 37.18132& 0.1691& 10.89& 11.42& 144.98& 265.50& 59.31& 95.83& 180.26& 1.66& 0.83& 1.68& -0.49& comp \\
GMBCGJ166.59455+09.23113& &166.59450& 9.23113& 0.2220& 10.87& 11.36& 67.82& 111.33& 44.42& 39.74& 205.95& 1.28& 1.86& 1.72& -0.01& comp \\
GMBCGJ167.06517+23.19665& &167.06520& 23.19666& 0.2223& 10.97& 11.47& 42.43& 58.70& 16.16& 20.86& 124.49& 0.68& 1.13&  1.92& -3.41& comp \\
GMBCGJ169.05610+57.02011& &169.05611& 57.02011& 0.1624& 10.71& 11.24& 270.35& 445.03& 38.74& 160.52& 244.49& 2.55& 1.09&  1.61& 0.90& comp \\
GMBCGJ176.68298+51.20676& &176.68300& 51.20676& 0.2792& 10.99& 11.54& 37.87& 47.64& 19.16& 17.69& 79.00& 0.92& 1.17&  1.90& 0.34& comp \\
GMBCGJ178.31993+23.51160& &178.31990& 23.51162& 0.2580& 11.05& 11.55& 401.35& 559.53& 221.72& 202.52& 1356.89& 9.03& 18.39&  1.56& 1.34& comp \\
GMBCGJ181.32281+33.82819& &181.32280& 33.82819& 0.1834& 10.66& 11.18& 34.70& 61.02& 20.47& 21.74& 42.38& 0.46& 0.20& 1.65& 0.26& comp \\
GMBCGJ183.10487+46.03118& &183.10490& 46.03118& 0.3433& 10.92& 11.54& 28.00& 52.30& 9.65& 18.49& 45.66& 1.63& 1.09&   1.97& 0.24& comp \\
GMBCGJ187.68381+36.55121& &187.68381& 36.55122& 0.1173& 10.68& 11.30& 352.08& 713.68& 67.96& 259.30& 405.45& 2.01& 0.88&   1.58& 0.80& sb \\
GMBCGJ187.85496+19.52685& &187.85500& 19.52687& 0.3289& 11.07& 11.68& 78.87& 123.96& 34.70& 44.02& 209.49& 3.50& 4.84&  2.10& -1.40& comp \\
GMBCGJ188.24063+42.90074& &188.24060& 42.90074& 0.1681& 11.06& 11.59& 805.24& 1249.19& 548.01& 472.21& 4424.96& 7.71& 23.11&  1.77& -0.99& comp \\
GMBCGJ191.18836+33.24089& &191.18840& 33.24089& 0.1298& 10.48& 11.12& 33.32& 45.77& 11.99& 17.31& 65.28& 0.16& 0.15&  1.94& -1.16& comp \\
GMBCGJ191.58395+32.91053& &191.58400& 32.91054& 0.1334& 10.74& 11.43& 103.01& 158.65& 73.53& 55.93& 244.64& 0.59& 0.66&  1.97& -2.44& comp \\
GMBCGJ192.83454+48.35832& &192.83450& 48.35832& 0.1479& 10.68& 11.25& 334.59& 726.67& 185.38& 260.61& 823.62& 3.39& 3.07&  1.41& 1.22& comp \\
GMBCGJ193.65889+07.03035& &193.65891& 7.03036& 0.3444& 11.10& 11.57& 18.42& 30.32& 6.43& 10.72& 22.59& 0.95& 0.51&  1.56& -0.89& comp \\
GMBCGJ201.34167+03.98025& &201.34171& 3.98024& 0.2548& 11.04& 11.64& 277.14& 322.76& 100.99& 116.48& 337.03& 5.06& 4.22&  1.63& -0.11& comp \\
GMBCGJ204.82740+33.85023& &204.82739& 33.85023& 0.2543& 10.96& 11.48& 106.13& 331.40& 94.69& 120.34& 412.61&  5.17& 5.23& 1.62& 1.74& sb \\
GMBCGJ210.45775+16.50052& &210.45770& 16.50051& 0.2194& 11.06& 11.52& 250.91& 308.32& 43.52& 109.59& 499.36& 3.45& 4.64&  1.81& -2.16& comp \\
GMBCGJ210.77165+10.05610& &210.77170& 10.05609& 0.1202& 10.50& 11.13& 44.78& 53.20& 21.54& 18.62& 67.36& 0.16& 0.12&  1.77& -1.00& comp \\
GMBCGJ215.06365+01.57007& &215.06360& 1.57007& 0.1905& 10.90& 11.36& 81.38& 113.22& 34.62& 40.03& 114.32& 0.92& 0.68&  1.74& -1.24& comp \\
GMBCGJ215.74059+27.34423& &215.74060& 27.34423& 0.1596& 11.00& 11.45& 226.44& 325.50& 102.67& 118.82& 998.77& 1.79& 4.55&  1.95& -2.61& comp \\
GMBCGJ216.45610+10.90254& &216.45610& 10.90254& 0.2400& 11.03& 11.55& 33.99& 35.42& 14.32& 16.68& 37.70& 0.48& 0.35&   1.84& -0.32& comp \\
GMBCGJ221.85842+08.47364$^{b}$& &221.85840& 8.47365& 0.3755& 11.27& 11.80& 1230.13& 1595.09& 272.50& 568.13& 2390.22& 61.30& 78.52&   1.25& 3.88& comp \\
GMBCGJ224.57182+50.96220& &224.57179& 50.96219& 0.2726& 10.90& 11.41& 41.29& 89.67& 46.08& 31.89& 223.36& 1.64& 3.29&   2.01& 0.17& comp \\
GMBCGJ224.62785+58.03770& &224.62790& 58.03770& 0.2819& 11.14& 11.66& 78.40& 106.68& 43.70& 38.33& 235.00& 2.11& 3.76&  1.79& -1.35& comp \\
GMBCGJ225.03764+21.65720& &225.03760& 21.65720& 0.1503& 10.91& 11.31& 676.11& 1909.93& 163.11& 707.53& 1301.82& 9.23& 5.19&   1.37& 3.68& sb \\
GMBCGJ229.56532+58.68779& &229.56531& 58.68780& 0.1877& 10.85& 11.48& 41.92& 52.55& 18.47& 19.11& 49.94& 0.41& 0.26&   1.83& -2.14& comp \\
GMBCGJ229.95329+26.36325& &229.95329& 26.36326& 0.1206& 10.64& 11.10& 210.19& 477.62& 55.02& 172.28& 197.96& 1.43& 0.42&  1.51& 0.82& sb \\
GMBCGJ230.61689+05.75409& &230.61690& 5.75410& 0.1825& 11.03& 11.38& 78.57& 101.46& 33.29& 36.05& 223.87& 0.75& 1.31& 2.02& -3.74& comp \\
GMBCGJ235.33215+18.81746& &235.33220& 18.81746& 0.3052& 11.30& 11.78& 150.07& 164.98& 38.05& 58.91& 294.47& 3.91& 5.78&  1.92& -2.44& comp \\
GMBCGJ241.00574+29.88749& &241.00571& 29.88749& 0.2927& 11.09& 11.61& 21.36& 43.40& 16.13& 15.98& 36.40& 0.93& 0.54&  1.74& 0.30& comp \\
GMBCGJ245.18841+26.31922& &245.18840& 26.31922& 0.2267& 10.95& 11.55& 150.90& 200.61& 99.14& 73.31& 834.61&  2.41& 8.38&  2.10& -3.14& comp \\
GMBCGJ245.35810+24.62761& &245.35809& 24.62761& 0.1878& 10.75& 11.32& 84.67& 118.88& 35.04& 42.58& 111.64& 0.94& 0.64&  1.65& -0.59& comp \\
GMBCGJ251.06514+45.86376& &251.06509& 45.86376& 0.1547& 10.85& 11.35& 46.33& 55.81& 23.06& 19.47& 80.21& 0.29& 0.28&  1.98& -1.10& comp \\
GMBCGJ258.86623+56.28854& &258.86621& 56.28856& 0.2903& 10.94& 11.36& 230.54& 383.25& 56.37& 137.95& 200.20& 8.10& 3.36&  1.32& 1.22& comp \\
GMBCGJ259.71676+31.73790& &259.71680& 31.73790& 0.1890& 10.42& 11.11& 49.74& 51.51& 11.21& 18.89& 59.51& 0.41& 0.34&  1.67& -0.66& comp \\
GMBCGJ315.89493+09.12085& &315.89490& 9.12084& 0.1458& 10.86& 11.36& 845.36& 1471.22& 526.45& 525.97& 2929.53& 6.65& 11.01&   1.94& -0.85& comp \\
GMBCGJ355.27875+00.30927$^{b}$& RX J2341.1+0018 &355.27869& 0.30927& 0.2768& 10.80& 11.58& 2504.19& 2947.64& 824.03& 1077.83& 6110.38& 55.84& 98.99& 1.28& 7.70& comp \\

\hline
\end{tabular}
\end{center}
{Note: \\
Col:(1) SDSS-WHL Cluster Name or GMBCG Cluster Name \\
Col:(2) 11 known X-ray luminous clusters out of 13 candidates. \\
Col:(3) BCG R.A.(J2000.0) in units of degrees. \\
Col:(4) BCG Dec.(J2000.0) in units of degrees. \\
Col:(5) The spectroscopic redshift of the BCG. \\
Col:(6) Logarithm of  the stellar mass inside the fiber aperture, in units of $M_{\odot}$. \\
Col:(7) Logarithm of the total stellar mass, in units of $M_{\odot}$. \\
Col:(8) The flux of ${\rm [N II]\lambda6584 }$ line in units of ${\rm 10^{-17} erg\, s^{-1} cm^{-2}}$. \\
Col:(9) The flux of ${\rm H_{\alpha}        }$ line in units of ${\rm 10^{-17} erg\, s^{-1} cm^{-2}}$. \\
Col:(10) The flux of ${\rm [O III]\lambda5007}$ line in units of ${\rm 10^{-17} erg\, s^{-1} cm^{-2}}$. \\
Col:(11) The flux of ${\rm H_{\beta}         }$ line in units of ${\rm 10^{-17} erg\, s^{-1} cm^{-2}}$. \\
Col:(12) The flux of ${\rm [O II]\lambda3727 }$ line in units of ${\rm 10^{-17} erg\, s^{-1} cm^{-2}}$. \\
Col:(13) The derived SFR($\rm H_{\alpha}$) by the ${\rm H_{\alpha}}$ emission line, in units of $M_\odot/\mathrm{yr}$. \\
Col:(14) The derived SFR($\rm [O II]$) by the ${\rm [O II]}$ emission line, in units of $M_\odot/\mathrm{yr}$. \\
Col:(15) The amplitude of the 4000 Balmer break, $D_n$(4000). \\
Col:(16) The absorption line index ${\rm H}{\delta}_A$. \\
Col:(17) Galaxy type classified by the BPT diagnosic diagrams: 'sb'$-$starburst, 'comp'$-$composite. \\
The SDSS-WHL objects marked with `a' are also in the GMBCG catalogue. \\
The objects marked with `b' are the identified 13 candidates of X-ray luminous clusters. \\ 
The flux of all emission lines used in this paper have been applied the dust correcton.
}
\end{minipage}
\end{table*}

\section{Data Analysis}

\subsection{SFR estimates \label{sec:assembly}}

It has been known that \Ha~emission is sensitive to the most recent star formation. The SFR based on \Ha~emission is an 
indicator of the nearly instantaneous SFR since it is produced by ionization by the hottest and youngest stars. 
We derive the SFRs of our target BCGs by the \Ha~line following \citet[][]{kennicutt98}
\begin{equation}
\mathrm{SFR(H_{\alpha})}=7.9\times10^{-42} L_\mathrm{H_{\alpha}}\,M_{\odot}\rm{yr^{-1}}
,
\end{equation}
where $L_\mathrm{H_{\alpha}}$ is the extinction-corrected luminosity of \Ha~emission in units of
$10^{42}\ \rm{ergs\ s^{-1}}$.
The derived SFRs(\Ha) range from 0.16 $M_\odot/\mathrm{yr}$ to 129.9 $M_\odot/\mathrm{yr}$, 
with an average value of 7.7 $M_\odot/\mathrm{yr}$.

\begin{figure}
\includegraphics[angle=0,width=0.47\textwidth]{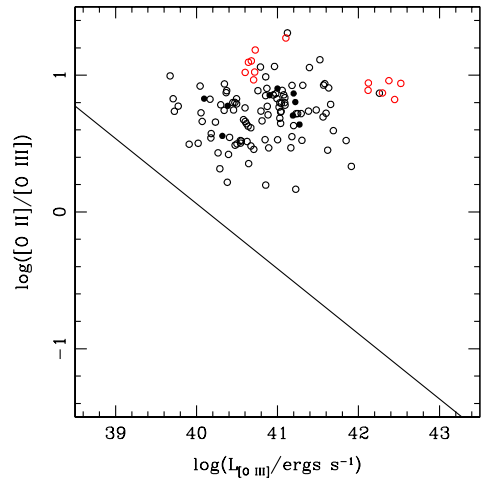}
\caption{ The line ratio [O II]/[O III] versus the
luminosity of [O III]. The solid line shows the least-square
regression for type I (narrow line) AGNs given by
\citet[][]{khi06}. The symbols are the same as in Figure~\ref{fig:diagnose}.
Our targets clearly show enhanced [O II]/[O III] line ratios relative to the \citet[][]{khi06} line. } \label{fig:kim}
\end{figure}

\begin{figure}
\includegraphics[angle=0,width=0.45\textwidth]{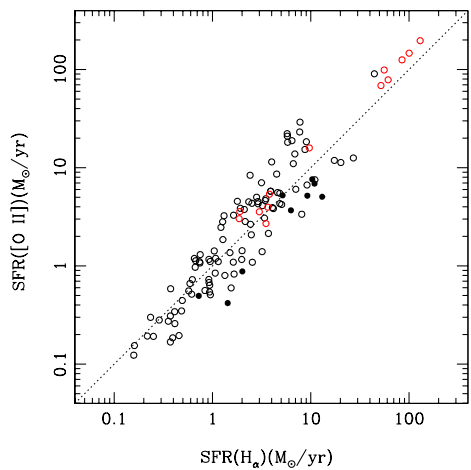}
\caption{
The comparison between the estimated SFR(\Ha) from the \Ha~line and SFR([O II]) from the [O II] line.
The symbols are the same as in Figure~\ref{fig:diagnose}.
It shows that these two estimates are roughly consistent with each other. 
} \label{fig:sfr-contrast}
\end{figure}

We also estimate their SFRs by the [OII]$\lambda$3727 line and make a comparison with SFRs(\Ha).
In order to estimate the contributions by AGN on the [O II] emission, we
follow \citet[][]{ww08} to investigate the luminosity of [O III]
as a function of the line ratio of [O II]/[O III] for our BCGs (see 
Figure~\ref{fig:kim}). The symbols are the same as in Figure~\ref{fig:diagnose} except
that BCGs in X-ray luminous clusters are shown with red symbols. 
Notice that hereafter we do not show the overlapping objects in the GMBCG catalog in the total sample.
It is clear that our composite BCGs show enhanced [O II]/[O III] ratios just like 
purely star-forming BCGs as these two types of objects reside in the same region. 

We can estimate the [O II] luminosities emitted from the HII regions by assuming the 
enhanced [O II]/[O III] ratios are caused by star formations \citep[][]{khi06}. 
The mean value of the [O II]/[O III] ratios is $\sim$ 7.0 for our BCGs. 
Given that the average [O III] luminosity $L_\mathrm{[O III]}$
$\sim$ 7.8 $\times$ 10$^{40}$ ergs s$^{-1}$, the [O II]/[O III]
ratio predicted by the regression line (see Figure~\ref{fig:kim})
given by \citet[][]{khi06} is only $\sim$ 0.43. It means that on average
$\sim$ 92\% of the [O II] emission  for our BCGs can be attributed to star formation. 

We use the recent calibration of \citet[][]{kgj04}
\begin{equation}
\mathrm{SFR(O~II)}=7.9\times\frac{C_1{\times}L_{\mathrm{[O~II],42}}}{16.73-1.75[\log(\mathrm{O/H})+12]}M_{\odot}\,\rm{yr^{-1}}
\end{equation}
to estimate the SFR(O II), where $L_\mathrm{[O II],42}$ is the
extinction-corrected luminosity of [O II] emission in units of
$10^{42}\ \rm{ergs\ s^{-1}}$. $C_1$ is the correction factor due to the enhanced [O II]/[O III] ratio. The metallicity is fixed to be $\log
(\rm{O/H})+12=8.9$, corresponding to the solar value. The
derived SFRs(O II) ranges from 0.12 $M_\odot/\mathrm{yr}$ to
196.9 $M_\odot/\mathrm{yr}$, with an average  value of 10.9
$M_\odot/\mathrm{yr}$.

\begin{figure*}
\includegraphics[angle=0,width=0.75\textwidth]{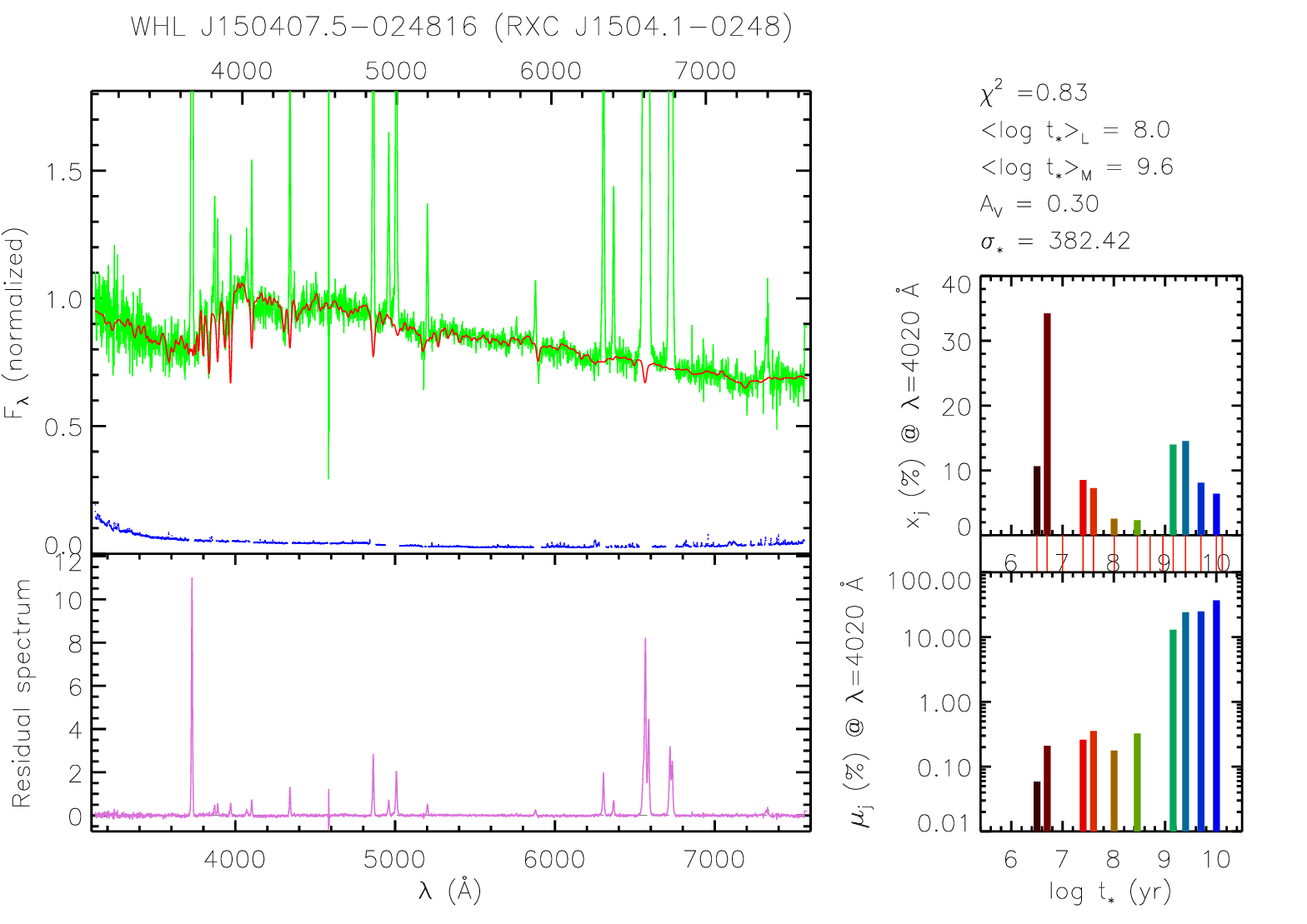}
\caption{
The spectral synthesis of an example BCG, WHLJ150407.5-024816 (RXC J1504.1-0248).
The observed spectrum $O_\lambda$ (green), model spectrum $M_\lambda$ (red) and
error spectrum (blue) of $O_\lambda$ 
are shown in the top left panel, respectively. The residual spectrum $E_\lambda$ (purple) is shown in the bottom left panel.
The flux intensities in the left two panels are both normalised at 4020{\rm \AA}\, by $4.5\times10^{-16}\,  {\rm ergs\ s^{-1}\ cm^{-2}}$.
The light and mass weighted stellar population fractions, $x_{\rm j}$ and  $\mu_{\rm j}$, are shown in the top right and bottom right panels,
respectively. Several derived quantities (see text for details) from the fitting are shown at the top right corner.
}
\label{fig:fitting}
\end{figure*}

The comparison between the SFR(\Ha) and SFR([O II]) for our BCGs is shown in Figure~\ref{fig:sfr-contrast}, 
which shows that these two estimates are consistent with each other and most of the [OII]$\lambda$3727 line emission 
in star-forming BCGs can be attributed to star formation.
It should be noted that our SFR estimates are from the SDSS spectra within 3$''$ fiber diameter, which corresponds to 
an average size of $\sim$ 10 kpc for our BCGs at $0.1<z<0.4$. It has been known that the majority of the line emissions in BCGs 
may be contained within this aperture \citep[e.g.,][]{hcf07}. We thus do not correct the aperture effect for our BCGs, 
as in previous studies \citep[e.g.,][]{cae+99,obp+08}. 

\subsection{Spectral synthesis}

\begin{figure*}
\includegraphics[angle=0,width=0.95\textwidth]{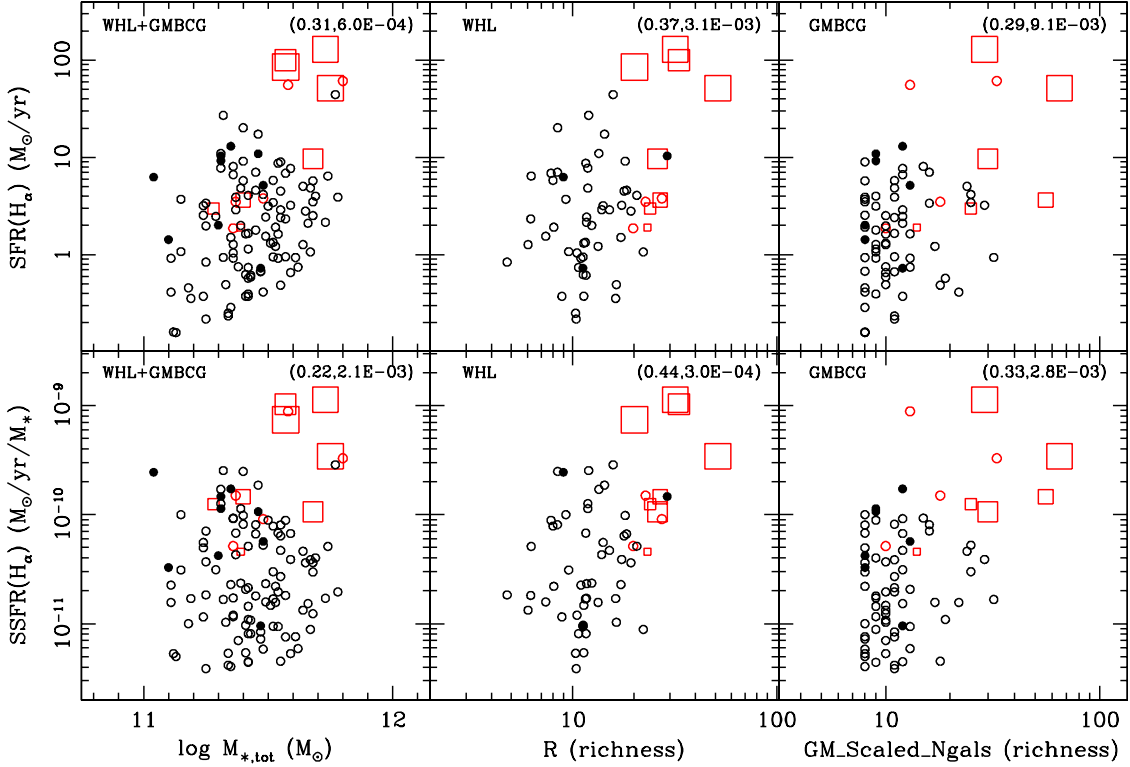}
\caption{ The derived total SFR and specific SFR (SSFR) by the \Ha~line versus the total stellar mass ($\rm log {M_{\rm \ast,tot}}$)
and cluster richness, respectively. The symbols are the same as in Figure~\ref{fig:diagnose}, except that 8 BCGs in known cooling flow clusters
are shown as red boxes. The size of each box is inversely proportional to its cooling time ($\rm t_{cool}$) (see Sec. 4.4).
At the top right corner of each panel, we show
the correlation coefficient and corresponding significance level for the null
hypothesis of no correlation as given by the Spearman-Rank order test.
}
\label{fig:environment2}
\end{figure*}

We use the spectral synthesis code \starlight\ \citep{cms+05} to derive stellar populations of our BCGs. 
\starlight\ fits the observed spectrum $O_\lambda$ with a model spectrum $M_\lambda$, 
which is made up of a pre-defined set of base spectra. It carries out the fitting with a simulated annealing plus 
Metropolis scheme to yield the minimum $\chi^2 = \sum_\lambda [(O_\lambda - M_\lambda) w_\lambda]^2$, where $w_\lambda^{-1}$ is 
the error in $O_\lambda$ at each wavelength. It models $M_\lambda$ by a combination
\begin{equation}
\label{Mlambda}
M_\lambda = M_\lambda(\vec{x}, A_V, v_\star, \sigma_\star) =
\sum_{j=1}^{N_\star} x_j \gamma_{j,\lambda} r_\lambda
\end{equation}
where $\gamma_{j,\lambda} \equiv b_{\lambda,j} \otimes G(v_\star, \sigma_\star)$,
$b_{\lambda,j} \equiv {B_{\lambda,j}\overwithdelims () B_{\lambda_{0},j}}$ is the normalised flux of the $j^{th}$ spectrum,
$B_{\lambda,j}$ is the $j^{th}$ component of base spectrum,
$B_{\lambda_{0},j}$ is the value of the $j^{th}$ base spectrum at the normalisation
wavelength $\lambda_0$, $G(v_\star, \sigma_\star)$ is the Gaussian distribution
centred at velocity $v_\star$ and  $\sigma_\star$ is the line-of-sight velocity dispersion, 
 $x_j$ is the fraction of flux due to component {\it j} at $\lambda_0$, 
$r_\lambda \equiv 10^{-0.4(A_\lambda-A_V)}$ is the global extinction term represented by $A_V$. The residual spectrum $E_\lambda$ 
including emission lines can be obtained by subtracting the model spectrum from the observed one as $E_\lambda = O_\lambda - M_\lambda$.

In this work, we take simple stellar populations (SSPs) from the
BC03 evolutionary synthesis models \citep{bc03} as our base spectra. 
We adopt the spectral templates with $N_\ast$$=42$ SSPs -- 3 metallicities (Z=0.2, 1 and 2.5 $Z_\odot$) 
and 14 ages (3, 5, 10, 25, 40, 100, 280, 500, 900 Myr and 1.4, 2.5, 5, 10, 13 Gyr), 
computed with the ``Padova 1994" evolutionary tracks \citep{abb+93,bfb+93,fbb+94a,fbb+94b,gbc+96} 
and the \citet{Chabrier03} initial mass function (IMF). We
follow \citet{mwg+10} to model the extinction using the 
dust extinction law given by Calzetti et al. (1994, 2000) and Calzetti (1997). We show a typical example of
spectral fitting for our BCGs in Figure~\ref{fig:fitting}. The top left panel 
shows the observed spectrum $O_\lambda$ (green) and the model $M_\lambda$ (red). 
The bottom left panel gives the residual spectrum $E_\lambda =O_\lambda - M_\lambda$ (purple). 
The light-weighted stellar population fractions $x_{\rm j}$ are shown in the top right panel. 
The mass-weighted population fractions $\mu_{\rm j}$ are shown in the bottom right panel.

\texttt{STARLIGHT} presents the current stellar mass and the
fraction of each stellar component. 
Following \citet{cms+05}, we can derive the mean ages of the stellar population weighted by the flux and stellar mass, respectively.
\begin{equation}
\label{logtl}
\langle {\rm log} t_\star \rangle_L = \sum _{j=1}^{N_\star} x_j \, {\rm log} \, t_j
\end{equation}
\noindent where $x_j$ is the fraction of flux contributed by certain SSP and
\begin{equation}
\label{logtm}
\langle {\rm log} t_\star \rangle_M = \sum _{j=1}^{N_\star} \mu_j \,{\rm log}\, t_j
\end{equation}
\noindent where $\mu_j$ is the fraction of stellar mass contributed by each SSP.
The results of the population synthesis are presented in \S4.2.

\section{Results} \label{sec:results}

\subsection{Dependence on stellar mass and environment} \label{sec:env}

\begin{table*}
\footnotesize
\setlength{\tabcolsep}{0.025in}
\caption{ The normalised percentage of flux and mass contributed on average by
stellar populations of different age for the whole optically-selected 120 BCGs with SF,
13 BCGs with SF in X-ray luminous clusters, 200 optically-selected quiescent BCGs,
and 116 quiescent (without distinct emission-lines) BCGs in X-ray luminous clusters,
respectively. The 1\,$\sigma$ (68.3\%) confidence interval for each average value is given in the parentheses.
}
\centering
\begin{tabular}{c|c|c|c|c}
\hline \multicolumn{1}{c|}{SSP} &
       \multicolumn{1}{c|}{SF (optical)}  &
       \multicolumn{1}{c|}{SF (X-ray)}  &
       \multicolumn{1}{c}{quiescent(optical)} &
       \multicolumn{1}{c}{quiescent(X-ray)}

\\
\hline
$f_{\rm burst}$ $\rm (t<0.1Gyr)$           & 7.63 (0.00,13.65)   & 18.37 (1.01,32.11)  & 1.38 (0.00,5.40) & 3.16 (0.00,7.12)\\
$f_{\rm young}$ $\rm (t <0.5Gyr)$          & 19.92 (3.91,39.10) & 39.77 (5.35,61.20) & 10.68 (0.00,17.34) & 9.24 (0.00,17.96)\\
$f_{\rm middle}$$\rm (0.5Gyr< t <2.5Gyr)$  & 24.31 (1.25,44.73) & 24.80 (6.12,50.04) & 20.53 (0.00,42.35) & 22.44 (0.00,36.54)\\
$f_{\rm old}$$\rm (t>2.5Gyr)$              & 55.77 (22.75,77.05) & 35.44 (17.22,58.56) & 68.79 (42.92,81.18) & 68.32 (48.23,88.91) \\
\hline
$m_{\rm burst}$ $\rm (t<0.1Gyr)$           & 0.04 (0.00,0.06)   & 0.17 (0.00,0.23)  & 0.002 (0.00,0.005) & 0.012 (0.00,0.016)\\
$m_{\rm young}$ $\rm (t <0.5Gyr)$          & 0.51 (0.00,1.87)   & 1.68 (0.00,2.97)  & 0.21 (0.00,0.59)  & 0.13 (0.00,0.34) \\
$m_{\rm middle}$$\rm (0.5Gyr< t <2.5Gyr)$  & 11.73 (0.00,28.36)  & 19.07 (2.00,40.12) & 4.65(0.00,9.60) & 5.72 (0.00,12.21)\\
$m_{\rm old}$$\rm (t>2.5Gyr)$              & 87.76 (57.35,93.06) & 79.25 (43.23,97.25)  & 95.14(88.20,97.60) & 94.14 (84.06,97.01)\\
\hline
\end{tabular}
\label{Table2}
\end{table*}

The stellar mass for each BCG inside the fiber aperture has been estimated from
our spectral synthesis, which are roughly consistent 
with that obtained by fits on the photometry by the MPA/JHU team. 
The specific SFR (inside the fiber aperture) for our BCGs can thus be derived.
The total stellar mass can be obtained by multiplying the factor, $C_2 \equiv 10^{-0.4(m_{\rm petro}-m_{\rm fiber})}$, 
between the fiber magnitudes and total (Petrosian) magnitudes. 
We follow the method of MPA/JHU team to take the correction factor averaged over the five SDSS bands, weighted by  $1/{\rm \Delta {C_2}^2}$ , where $\rm \Delta {C_2}$ 
is the error in the correction factor $C_2$ (estimated from the errors in the photometry).
The estimated SFR and specific SFR by the \Ha~line versus the galaxy total stellar mass ($\rm \log\, {M_{\rm \ast,tot}}$) and the cluster richness 
for our BCGs are shown in Figure~\ref{fig:environment2} (the relations are similar if the SFR estimated by the [O II] line is used), respectively. 
The symbols are the same as in Figure~\ref{fig:diagnose}, except that 8 BCGs in known cooling flow clusters
are shown as red boxes. The size of each box is inversely proportional to its cooling time ($\rm t_{cool}$). 
Notice that the rest 5 BCGs in X-ray luminous sample are also likely to be in cooling flow clusters (see \S4.4 for discussions).
We perform the Spearman-Rank order correlation test for each relation. The corresponding correlation coefficient 
and the significance level of the null hypothesis that there is no correlation are given in each panel of Figure~\ref{fig:environment2}. 
It can be seen that there is an obvious trend that more massive BCGs with SF and those in richer clusters tend to have higher SFR and specific SFR, but with large scatters.
BCGs with SF in X-ray luminous clusters are often located in the densest environment and have the highest SFR and specific SFR, 
which shows they appear to be forming stars at a higher rate. 
BCGs with SF in cooling flow clusters (red boxes) usually have the most active star formation (also see \S4.4).

\subsection{Star formation history} \label{sec:sfh}

We investigate the star formation history of these BCGs with SF through spectral synthesis method. 
We combine 42 SSPs into three ages: young-, middle- and old-age stellar populations. 
The young-age stellar population includes the SSPs with age less than 0.5 Gyr, 
the old-age population is  SSPs with age larger than 2.5 Gyr, 
and the intermediate-age population is the SSPs between them. We also show a burst population, 
defined as  SSPs with age less than 0.1 Gyr, connected with the most recent star formation activity. 
We make a direct comparison between these BCGs with SF and a randomly-selected sample 
of 200 quiescent BCGs (without emission lines of both \Ha~and [O II], i.e. sources 
with the EWs of \Ha~and [O II] $\leq 0$ excluded at the 95.4\% confidence level)
in MPA/JHU catalogues.       
%
These two samples are matched in total stellar mass. We have shown that the majority ($\sim80\%$) of BCGs in two X-ray luminous samples are not classified 
as emission-line BCGs. They have also no significant emission lines (`quiescent'). 
We make another control sample with 116 objects.
The normalised fractions of their stellar populations are listed in Table 2, which provide a coarse star formation history of these BCGs. 

\begin{figure}
\includegraphics[angle=0,width=0.44\textwidth]{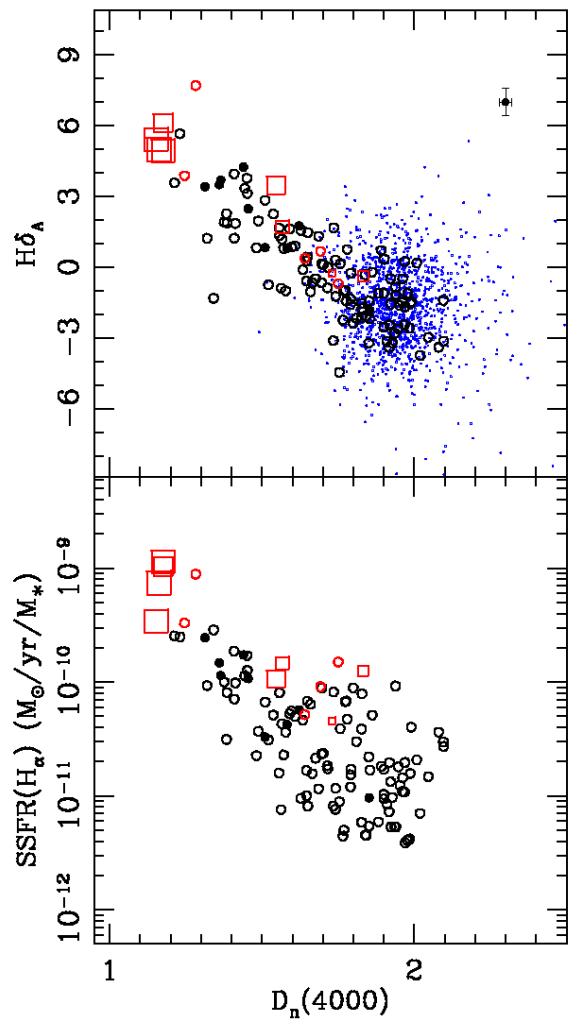}
\caption{Top panel: the Balmer line absorption index, $\Hdelta_A$ plotted
against the 4000\AA~break strength, $D_n(4000)$ for the  BCGs with SF and quiescent BCGs (blue dots),
respectively.
Bottom panel: specific SFR (SSFR) versus the 4000\AA~break strength, $D_n(4000)$ for the  BCGs with SF.
The symbols are the same as in Figure~\ref{fig:diagnose}. The BCGs in X-ray luminous clusters are shown with red symbols, in particular,  8 BCGs in known cooling flow clusters are shown as red boxes. The sizes of boxes
are inversely proportional to their cooling times ($\rm t_{cool}$).
} \label{fig:hdelta-d4000}
\end{figure}

The flux-weighted average population fraction is sensitive 
to the star formation activity. Nearly 20\% flux are from the young stellar population 
for the whole sample. This fraction can increase to $\sim$40\% for BCGs 
with SF in X-ray luminous clusters. A large fraction ($\sim$40\%) 
of the young stellar population is from the recent burst (with age $\rm t <0.1Gyr$). 
BCGs with SF and quiescent BCGs have comparable fraction of intermediate-age stellar population, which 
indicates that SF activities in BCGs contribute mainly to the fraction of young stellar population and 
the timescale of SF activity is short. However, the fractions of stellar mass 
in different age bins are significantly different from the flux-weighted ones. 
The majority of stellar mass of BCGs with SF is contributed by the old population. 
The stellar mass of the young stellar population is small ($\sim 0.5\%$ on average), 
which is consistent with the result of \citet[][]{pkb+09}  obtained from a smaller sample. 
It shows that the stellar population will not have significant differences with that of normal BCGs when their
star formation is quenched. The stellar population of BCGs with SF are still predominantly old, not very different with the quiescent (normal) BCGs. 

Notice that about $\sim 12\%$ of stars formed within the last 2.5 Gyr or so for our BCGs with SF (the first column in Table 2). The derived stellar mass inside the fiber aperture is $8.2\times10^{10}$$\rm M_\odot$ on average. If the typical SFR ($\sim 7.7 M_\odot\,{\rm yr}^{-1}$) in bursts is the same as the current one, then the total star formation duration will be 1.3 Gyr. This is consistent with a scenario that the rejuvenation SF activities in these BCGs may be sporadic \citep[e.g.,][]{sts+07}. 
In any case, it should be emphasized that the scatters in these estimates are quite high (see Table 2).

It has been shown by \citet[][]{khw+03a} that the plane
defined by the 4000\AA~break strength, $D_n(4000)$, and Balmer
line absorption index, $\Hdelta_A$, is also a powerful diagnostic for
the star formation history of galaxies. We show the $\Hdelta_A$ absorption index as
a function of $D_n(4000)$ for these BCGs with SF in the top panel 
of Figure~\ref{fig:hdelta-d4000}, and compare with quiescent 
BCGs (blue dots).  We also show the relation of the specific SFR versus $D_n(4000)$ 
for our BCGs with SF in the bottom panel of Figure~\ref{fig:hdelta-d4000}. 
The values of $D_n(4000)$ and $\Hdelta_A$ are taken from the MPA/JHU catalog. 
As can be seen, BCGs with more active SF activities (higher specific SFR) tend to have lower $D_n(4000)$ 
values and stronger $\Hdelta_A$ absorption than quiescent BCGs. 
It means they have a higher fraction of young stars, and are more likely 
to be experiencing sporadic star formation events at the present day \citep[][]{khw+03a}.
This analysis is consistent with our results obtained through spectral synthesis method.

\subsection{SF activity \& cooling flow} 

\begin{figure*}
\includegraphics[angle=0,width=0.9\textwidth]{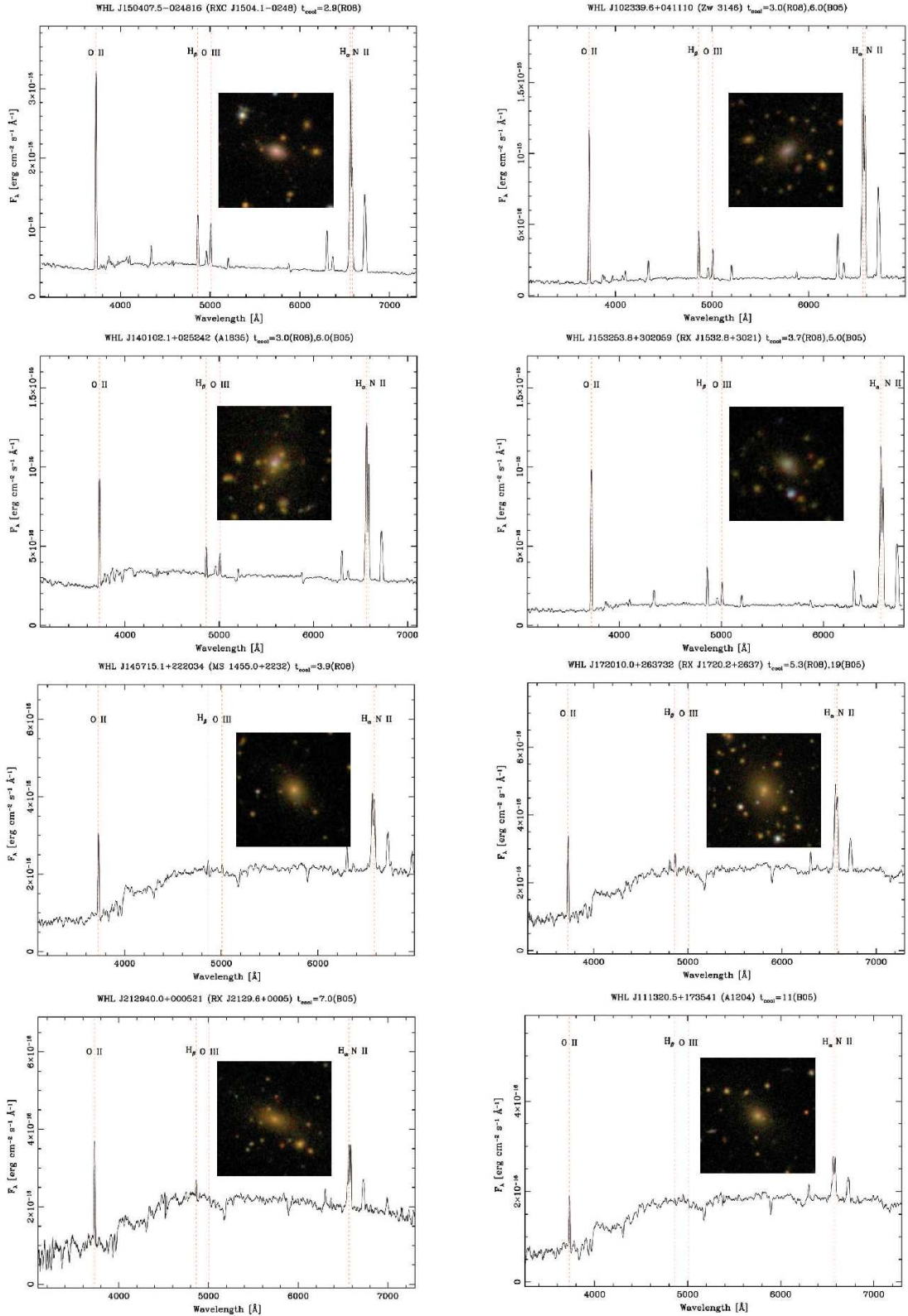}
\caption{ Spectra and colour images of the  eight known early-type BCGs with significant ongoing SF in cooling flow clusters. 
Each spectrum is shifted to the rest-frame wavelength, corrected for the Galactic extinction, and smoothed 
using a 15 \AA~box. The size of each colour image corresponds to 200\,kpc by 200\,kpc. The objects are ordered with increasing 
cooling time ($\rm t_{cool}$) in units of $10^8 yr$,  taken from \citet[][R08]{rmn08} and/or \citet[][B05]{bfs+05}.  
}
\label{fig:cooling}
\end{figure*}

It has been shown that active star formation in BCGs may be connected with the cooling flow of intracluster medium (ICM) in many X-ray 
clusters  \citep[e.g.,][and reference therein]{rmn08}. There are 11 known BCGs in our 13 targets in X-ray luminous clusters. 
We collect their information from the literature, and find that 8 out of 11 ($\sim73\%$) have been identified to be 
in cooling flow clusters \citep[e.g.,][]{df08,rmn08}. 
The rest-frame optical spectra and colour images of BCGs in these eight clusters (RXC J1504.1-0248, Zw 3146, A1835, 
RX J1532.8+3021, MS 1455.0+2232, RX J1720.2+2637, RX J2129.6+0005, A1204) are shown 
in Figure~\ref{fig:cooling}, ordered with increasing cooling time of ICM. Their cooling time $\rm t_{cool}$ 
are obtained from \citet[][]{rmn08} and/or \citet[][]{bfs+05}, which are marked in each panel of Figure~\ref{fig:cooling}. 
Although not all objects have measured cooling time and the derived values are somewhat different for the same source from the two studies, 
it is still apparent in Figure~\ref{fig:cooling} that the four BCGs with shortest cooling times have bluer colours, 
which is consistent with the result of \citet[][]{rmn08} that bluer BCGs reside in clusters with shorter cooling times. 
These eight BCGs in cooling flow clusters are shown with red boxes in the plot of specific SFR versus $D_n(4000)$ (the bottom panel of Figure~\ref{fig:hdelta-d4000}), 
with the size of each box inversely proportional to the cooling time. It can be seen that BCGs in cooling flow clusters 
usually have more active SF activities (higher specific SFR): the four BCGs (RXC J1504.1-0248, Zw 3146, A1835, RX J1532.8+3021) 
with the shortest $\rm t_{cool}$ have the most active SF activities. In fact, it has been shown two (Zw 3146 and A1835) of them can even 
be classified as luminous infrared galaxies (LIRGs) \citep[][]{emr+06}. This is a strong indicator that the cooling 
flow and star formation in these BCGs are connected. Figure~\ref{fig:cooling} also shows that BCGs in cooling flow clusters 
(in particular the four with the shortest $\rm t_{cool}$) usually have very flat optical spectra (smaller 4000\AA~break strength, see the bottom panel of Figure~\ref{fig:hdelta-d4000}).

\begin{figure*}
\includegraphics[angle=0,width=0.9\textwidth]{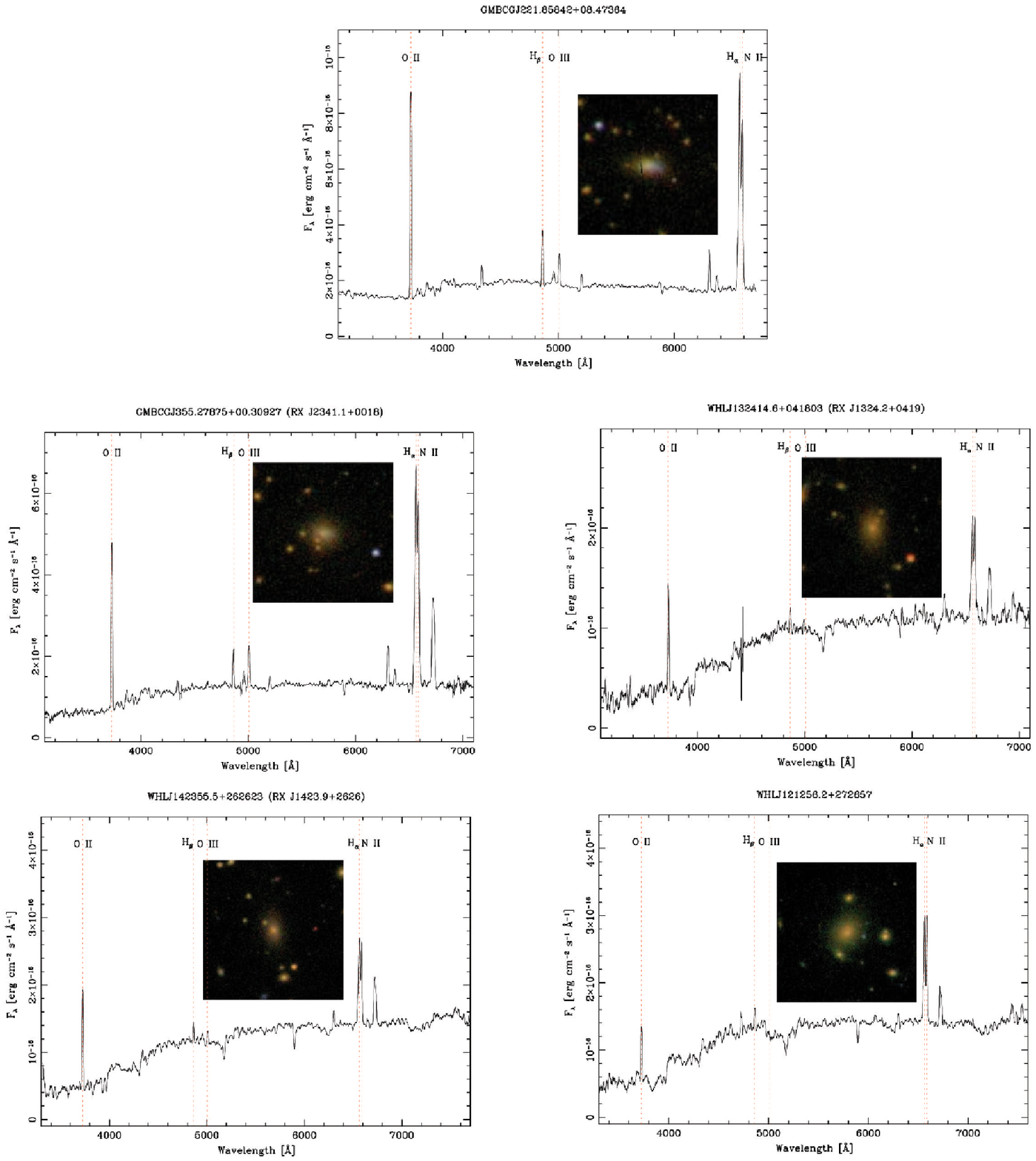}
\caption{ 
Spectra and colour images of the other 5 early-type BCGs with ongoing SF in 13 X-ray luminous clusters. 
Each spectrum is shifted to the rest-frame wavelength, corrected for the Galactic extinction, and smoothed
using a 15\,\AA~box. The size of each colour image corresponds to 200\,kpc by 200\,kpc.
}
\label{fig:restXray}
\end{figure*}

It should be noted that we do not know whether or not the remaining  5 targets in our X-ray luminous sample are also 
in cooling flow clusters. Their rest-frame spectra and colour images are shown in Figure~\ref{fig:restXray}.
It can be seen that their images and optical spectra are very similar to those of 8 BCGs in known cooling flow clusters, 
particularly for two GMBCG objects (GMBCG J221.85842+08.47364, J355.27875+00.30927). 
These two objects have extremely blue images and very flat optical spectra, with extremely high SFR \& SSFR 
(two red circles on the top of right panels of Figure~\ref{fig:environment2}). In any case, 
the majority (even 100\%) of BCGs with SF in X-ray luminous sample are likely to be in cooling flow clusters.

\section{Summary \& Discussion} \label{sec:discussion}

Only a few BCGs have been reported with ongoing star formation in previous studies and the majority of them are identified 
from X-ray cluster samples. In this paper, we identify a large sample of 120 early-type BCGs at $0.1<z<0.4$  from two large
optically-selected cluster catalogues of SDSS-WHL \citep[][]{WHL09} and GMBCG \citep[][]{hmk+10}. Their optical spectra 
show strong emission lines of both [O II]${\lambda}$3727 and \Ha, indicating significant ongoing star formation. 
This sample is not biased toward X-ray luminous clusters, and is thus more representative of this population.
We investigate their statistical properties and make a comparison with a control sample selected from X-ray luminous clusters. 
We also investigate their star formation history using stellar population synthesis models. The main results can be summarised as follows. 

\begin{enumerate}
\item
The incidence rates of emission-line BCGs and BCGs with SF in X-ray luminous clusters 
are almost one order of magnitude higher than those in optically-selected clusters.
\item 
More massive BCGs with SF in richer clusters tend to have higher SFR and specific SFR, 
which shows they appear to be forming stars at a higher rate. BCGs with SF in X-ray luminous clusters 
usually have more active SF activities.  
\item
The star formation history of BCGs with SF can be well described by a recent minor and short starburst superimposed 
on an old stellar component, with the recent episode of star formation contributing $<$ 1 percent of the total stellar mass ($\sim 0.5\%$ on average). 
The star formation history may be episodic, lasting a substantial fraction of the time in the last 2.5 Gyr (see Table 2 and section 4.2).
\item
BCGs with SF in cooling flow clusters usually have very flat optical spectrum and the most active SF activities. 
Star formation in these BCGs and cooling flow are correlated. 
\end{enumerate}

\begin{figure*}
\includegraphics[angle=0,width=0.95\textwidth]{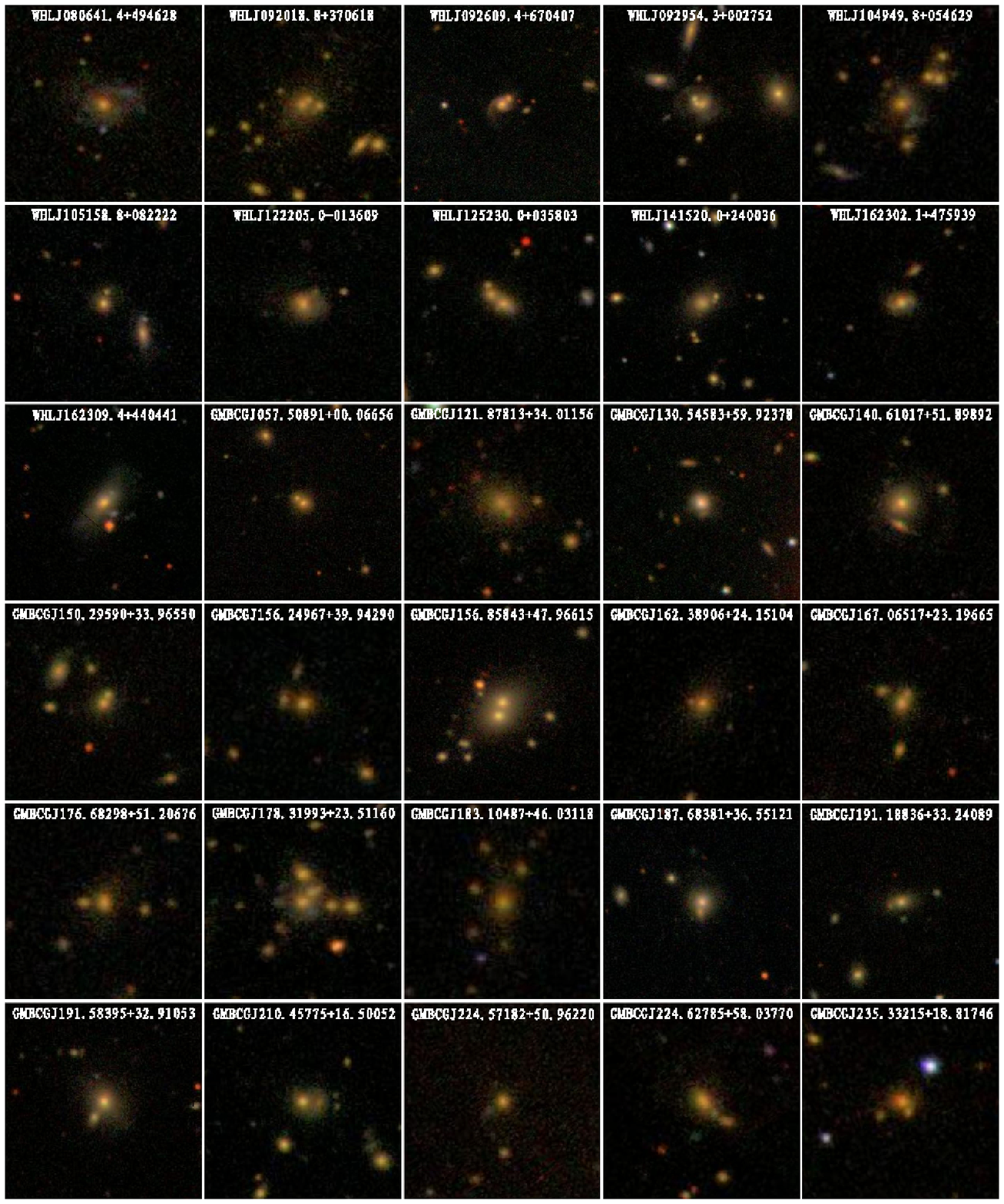}
\caption{ Colour images of 30 sample early-type BCGs with distinct merger features in non-X-ray luminous clusters.
The size of each colour image corresponds to 200\,kpc by 200\,kpc. 
}
\label{fig:minor-merger}
\end{figure*}

Although the short SF activity appears to be a rather common phenomenon during the evolution of BCGs, the source of the cold gas 
required to fuel the star formation is unclear. The correlation between SF activities in BCGs in cooling flow clusters 
and the gas cooling timescale \citep[][]{rmn08} suggests a clear (cooling) origin of the cold gas in these systems. 
However, the majority of our star-forming BCGs have less active SF activities. They either lie in non-cooling flow (non-X-ray luminous) clusters 
or may be below the ROSAT detection limit.
Recent theoretical studies often assume a very efficient form of AGN feedback, which may suppress the star formation completely. 
On the other hand, \citet[][]{bhb+08} suggest that AGN feedback may not fully compensate the energy lost via radiative cooling, 
allowing the gas to cool at a  reduced rate. It remains to be seen how universal this mode operates in BCGs and what kind of SFR it can sustain. 
Another possible and attractive mechanism is through the known galactic cannibalism that appears 
frequently in cluster environments due to dynamical friction. We indeed find a large fraction of sample BCGs in non-X-ray luminous clusters with distinct 
merger features (see 30 examples\footnote{Colour images and corresponding spectra for all 120 target BCGs are 
available at http:///www.jb.man.ac.uk/\~\,smao/liu.tar.gz} in Figure~\ref{fig:minor-merger}). 
If there is some remaining cold gas in the captured satellite galaxy, 
then the merger may supply fresh cold gas ($\sim$ few $10^8M_\odot$) that can trigger a new episode of star formation  \citep[][]{pfb+06}. 
It will be interesting to explore this issue further using our sample in a future study.

\section*{Acknowledgments}

We thank X. Y. Xia, Z. G. Deng, Z. L. Wen, Cheng Li, Lin Yan, J. Wang for useful discussions and comments, 
We acknowledge the anonymous referee for a constructive report that much improved the paper. 
This project is supported by the NSF of China 11103013. SM and XMM acknowledge the Chinese Academy of Sciences for financial support.

Funding for the creation and distribution of the SDSS Archive has
been provided by the Alfred P. Sloan Foundation, the Participating
Institutions, the National Aeronautics and Space Administration,
the National Science Foundation, the U.S. Department of Energy,
the Japanese Monbukagakusho, and the Max Planck Society. The SDSS
Web site is http://www.sdss.org/. The SDSS is managed by the
Astrophysical Research Consortium (ARC) for the Participating
Institutions. The Participating Institutions are The University of
Chicago, Fermilab, the Institute for Advanced Study, the Japan
Participation Group, The Johns Hopkins University, the Korean
Scientist Group, Los Alamos National Laboratory, the
Max-Planck-Institute for Astronomy (MPIA), the
Max-Planck-Institute for Astrophysics (MPA), New Mexico State
University, University of Pittsburgh, Princeton University, the
United States Naval Observatory, and the University of Washington.


\label{lastpage}
\end{document}